\documentclass[journal,onecolumn]{IEEEtran}
\usepackage[T1]{fontenc}
\usepackage[linesnumbered,ruled]{algorithm2e}
\usepackage{stfloats,color}
\usepackage{amsfonts}
\usepackage{amssymb}
\usepackage[cmex10]{amsmath}
\usepackage{graphicx}
\usepackage{setspace}
\usepackage{cite}
\usepackage{array}
\usepackage{subcaption}
\usepackage{times}
\usepackage{epsfig}
\usepackage{latexsym}
\usepackage{epstopdf}
\usepackage{verbatim}
\usepackage{units}
\usepackage{amsthm}
\usepackage{placeins}
\usepackage{afterpage}
\usepackage{dsfont}
\usepackage{soul}
\usepackage{multicol}
\usepackage{multirow}
\usepackage{mathtools}
\usepackage[cmintegrals]{newtxmath}
\newcolumntype{P}[1]{>{\centering\arraybackslash}p{#1}}
\newcolumntype{M}[1]{>{\centering\arraybackslash}m{#1}}

\newcommand{\defeq}{\ensuremath{\triangleq}}
\newtheorem{lemma}{Lemma}

\newtheorem{definition}{Definition}
\newtheorem{remark}{Remark}

\ifCLASSINFOpdf
\else
\fi
\begin{document}
%
\title{Uncoded Caching and Cross-level Coded Delivery \\ for Non-uniform File Popularity \\
}
%
%
%

\author{Emre~Ozfatura and
        Deniz~G{\"u}nd{\"u}z       
\thanks{Emre Ozfatura and Deniz G{\"u}nd{\"u}z are with the Information Processing and Communications Lab, Department of Electrical and Electronic Engineering, Imperial College London, London SW7 2AZ, UK. }
\thanks{This work was supported by EC H2020-MSCA-ITN-2015 project SCAVENGE under grant number 675891, and by the European Research Council project BEACON under grant number 677854.}
\thanks{A shorter version of this paper was presented at IEEE International Conference on Communications (ICC), in 20-24 May 2018.}}

%
%

\markboth{}%
{}
%



\maketitle

\begin{abstract}
Proactive  content caching at user devices and coded delivery is studied for a non-uniform file popularity distribution. A novel centralized uncoded caching and coded delivery scheme, called {\em cross-level coded delivery (CLCD)}, is proposed, which can be applied to large file libraries under non-uniform demands. In the CLCD scheme, the same sub-packetization is used for all the files in the library in order to prevent additional zero-padding in the delivery phase, and unlike the existing schemes in the literature, users requesting files from different popularity groups can still be served by the same multicast message in order to reduce the delivery rate. Simulation results indicate more than $10\%$ reduction in the average delivery rate for typical Zipf distribution parameter values. 
\end{abstract}


%
\IEEEpeerreviewmaketitle

\section{Introduction}\label{s:Intro}
Proactive caching is a well known strategy to reduce the network load and delivery latency in content delivery networks.  Recently, it has been shown that proactive caching and coded delivery, together, can further help to reduce the network load. The ``shared-link problem'', in which a server with a finite file library serves the demands of cache-equipped users over a shared error-free link is analyzed in \cite{CD.F1}, where the objective is to design a caching and delivery strategy to minimize the load on the shared link, often referred to as the ``delivery rate'' in the literature.The proposed solution  consists of two phases. In the {\em placement phase}, all the files in the library are divided into sub-files, and each user proactively caches a subset of these sub-files. In the {\em delivery phase}, the server multicasts XORed versions of the requested sub-files through the shared link, such that each  user can recover its demand by exploiting  both the cached sub-files and the multicast messages. In the scheme proposed in \cite{CD.F1}, the server controls the caching decisions of the users in a centralized manner. However, in \cite{CD.F2}, the authors have shown that  similar gains can still be achieved when the caching is performed in a random and decentralized manner under certain assumptions. It is shown in \cite{CD.F3} that the delivery rate can be further reduced with a coded delivery scheme that exploits the common requests. 
Most of the initial studies on coded caching and delivery  have been built upon certain simplifying unrealistic assumptions, such as  the presence of an error-free broadcast link, uniform user cache capacities, and uniform file demands. A plethora of works have since focused on trying to remove or relax these assumptions in order to bring  coded caching and delivery closer to reality. To this end, the case of non-uniform cache sizes is studied in \cite{CD.NCS1,CD.NCS2,CD.NCS3}, extension to unequal link rates is studied in \cite{CD.ULR1,CD.NCS2,CD.ULR3}, and  non-uniform file popularity is considered in \cite{CD.ND1,CD.ND2,CD.ND3,CD.ND4,CD.ND5,CD.ND6,CD.ND7}.\\
\indent In this work, we are interested in the case of non-uniform demands as this is commonly encountered in practice\cite{VS1,VS2,VS3,VS4}.It has been shown that by taking into account the statistics of the content popularity and by allocating the cache memory to the files in the library based on this statistics, communication load can be further reduced \cite{CD.ND1,CD.ND2,CD.ND3,CD.ND4,CD.ND5,CD.ND6,CD.ND7}. Most of the papers dealing with non-uniform demands consider  decentralized caching in the placement phase. In \cite{CD.ND1}, the authors proposed a {\em group-based memory sharing} approach for the placement phase, where the files in the library are grouped according to their popularities and the available cache memory at user devices is distributed among these groups. Then, in the delivery phase, the coded delivery scheme in \cite{CD.F2} is used to deliver the missing parts of the demands. In \cite{CD.ND2}, the authors introduced both a placement and a delivery strategy for non-uniform demands.  Optimization of the delivery phase is analyzed as an index coding problem and the multicast messages are constructed based on the {\em clique cover problem}, where each missing sub-file\footnote{We use the term {\em sub-file} to be consistent with the later part of the paper, while the papers studying the asymptotic performance of decentralized coded caching typically use the term "packet" or "bit".} (requested but not available in the user's cache) is considered as a vertex of a conflict graph, and two sub-files are delivered together in the same multicast message if  there is no edge between the corresponding vertices. The chromatic number of the conflict graph gives the number of required multicast messages, equivalently, the delivery rate normalized by the sub-file size. Due to the NP-hardness of the original graph coloring problem, the authors introduced a low-complexity suboptimal {\em greedy constrained coloring (GCC)} scheme. For the placement phase, similarly to \cite{CD.ND1}, a group-based memory sharing approach with only two groups is considered, where only the files in the first group are cached. They show that the proposed scheme is order-optimal when the popularity of the files follows a Zipf distribution. Results of \cite{CD.ND2} are further extended by \cite{CD.ND3} to show that group-based memory sharing with two groups\footnote{ In certain cases there might be an additional group consist of only a single file.} is order-optimal for any popularity distribution. The group-based memory sharing approach for non-uniform demands is also studied in \cite{CD.ND4},  where the users do not have caches, but instead have access to multiple cache-enabled access points and the shared link is between the server and the access points. All these works analyze the asymptotic performance (as the number of sub-files goes to infinity) of the decentralized coded caching approach.\\       
\indent Decentralized coded caching  with finite number of sub-files is recently studied in \cite{CD.ND7},\cite{CD.Dopt1} and \cite{CD.Dopt2} and  to provide an efficient low-complexity solution to the clique cover problem. Hence, as a common objective, the proposed methods in these works aim to minimize the delivery rate for a given placement strategy using heuristic approaches.\\
\indent For the centralized coded caching problem under a non-uniform demand distribution, a novel content placement strategy is introduced in \cite{CD.ND5} and \cite{CD.ND6} independently. In these papers, content placement is formulated as a linear optimization problem, where each file is divided into $K+1$ disjoint fragments, not necessarily of equal size. The $k$th fragment of each file is cached as in the placement phase of \cite{CD.F1} with parameter $t=k-1$. In the proposed approach, the size of each fragment of each file is represented by a variable in the linear optimization problem, and the solution of the optimization problem reveals how each file should be divided into $K+1$ fragments. We remark that the group-based memory sharing scheme of \cite{CD.ND1} is a special case of these approaches, where in each file group only one fragment has non-zero size. Since this approach considers the most general placement strategy, we call it {\em general memory sharing}, while we will refer to the original scheme in \cite{CD.F1} as the {\em naive memory sharing} scheme.
Note that, in the delivery phase, when a multicast message contains sub-files with different sizes, the size of the message is determined according to the largest sub-file in the message, and the smaller are zero-padded, which results in inefficiency.\\
\indent Hence, although the above scheme allows each file to be divided into different size fragments, when the file library is large, the proposed policy tends to divide the library into small number of groups, where, in each group, the size of the fragments are identical. This is because unequal fragment sizes among users requires zero-padding. In particular, we observe that, for a large file library, the proposed scheme in \cite{CD.ND6} often tends to classify the files into two groups according to their popularities, such that only the popular files are cached in an identical manner. We highlight here that without mitigating the sub-file mismatch issue, which induces zero padding, the advantage of optimization over fragment sizes is limited. Another limitation of the strategy proposed in \cite{CD.ND6} is its complexity.
The number variables in the optimization problem is $N(K+1)$. Although linear optimization can be solved in polynomial time, it may still be infeasible when the library size is large (in practical scenarios the number of files are often considered to be between $10^{4}-10^{6}$ \cite{VS2}).
\indent In the case of less skewed popularity distributions using more than two groups can be effective in reducing the delivery rate,  together with a more intelligent  coded delivery design that reduces or prevents zero-padding. This motivates the proposed {\em cross-level coded delivery (CLCD)} scheme, whose goal is to design a zero-padding-free coded delivery scheme for a given group structure. The zero-padding problem in the delivery phase of  centralized coded caching was previously studied in \cite{CD.ND7}, where a heuristic {\em ``bit borrowing ''} approach is introduced. The proposed algorithm greedily pairs the bits of the sub-files for a given placement strategy.\\ 
\indent In this paper, different from the aforementioned works, we introduce a novel centralized coded caching framework, where the delivery and the placement phases are optimized jointly. In the proposed CLCD strategy, the placement phase is limited to three file groups, formed according to popularities. Then, for each possible demand realization, i.e., the number of users requesting a file from each group, we obtain the optimal delivery scheme and the corresponding delivery rate. Finally, using this information we find the optimal group sizes based on the popularity of the files in the library. We will show that the proposed CLCD scheme  provides a noticable reduction in the average delivery rate, up to 10-20\%, compared to the state-of-the-art.\\
The rest of the paper is organized as follows. In Section \ref{s:SystemModel}, we introduce the system model and the problem definition. In Section \ref{s:CLCD}, we introduce the CLCD scheme and analyze it for a particular scenario, where we impose certain constraints on the placement phase. In Section \ref{s:clcd_ext}, we extend our analysis by lifting some constraints on the placement phase, and in Section \ref{s:sim}, we present numerical comparisons of the proposed CLCD scheme with the state-of-the-art. In Section \ref{s:clcd_ext2}, we present the most general form of the CLCD scheme without any constraints on the placement phase, and finally, we conclude the paper in Section \ref{s:conclusion}.

\section{System Model}\label{s:SystemModel}
We consider a single content server with a library of $N$ files, each of size $F$ bits, denoted by  $W_{1},\ldots,W_{N}$, serving $K$ users, each with a cache memory of capacity $MF$ bits. The users are connected to the server through a shared error-free link. We follow the two-phase model in \cite{CD.F1}. Caches are filled during the {\em placement phase} without the knowledge of particular user demands. User requests are revealed in the delivery phase, and are satisfied simultaneously.\\
\indent The request of user $k$ is denoted by $d_{k}$, $d_{k}\in \left[N \right]\defeq\left\{1,\ldots,N\right\}$, and the demand vector is denoted by $\mathbf{d}\defeq\left(d_{1},\ldots,d_{K}\right)$. The corresponding delivery rate $R(\mathbf{d})$ is defined as the total number of bits sent over the shared link, normalized by the file size. We assume that the user requests are independent of each other, and each user requests file $W_n$ with probability $p_n$, where $p_1 \geq p_2 \geq \cdots \geq p_N$, and $\sum_{n=1}^N p_n = 1$. Let $P(\mathbf{d})$ denote the probability of observing demand vector $\mathbf{d}$, where $P(\mathbf{d})=\Pi^{K}_{k=1} p_{d_{k}}$. We want to minimize the average delivery rate $\bar{R}$, defined as
\begin{equation}
\bar{R}\defeq \sum_{\mathbf{d}}P(\mathbf{d})R(\mathbf{d}).
\end{equation}
\indent Next, we  explain the uncoded caching and centralized coded delivery scheme, introduced in \cite{CD.F1}. We say that a file $W_n$ is  {\em cached at  level} $t$, if it is divided into ${K \choose t}$ non-overlapping sub-files of equal size, and each sub-file is cached by a distinct subset of $t$ users. Then, each sub-file can be identified by an index term $\mathcal{I}$, where $\mathcal{I}\subseteq \left[K\right]$ and $\vert \mathcal{I}\vert =t$, such that sub-file $W_{n,\mathcal{I}}$ is cached by users $k\in\mathcal{I}$. Following a placement phase in which all the files are cached at level $t$, as proposed in \cite{CD.F1}, in the delivery phase, for each subset $\mathcal{S}\subseteq \left[K\right]$, $\vert \mathcal{S} \vert =t+1$, all the requests of the users in $\mathcal{S}$ can be served simultaneously by multicasting 
\begin{equation}\label{coddel}
\bigoplus_{s\in \mathcal{S}}W_{d_{s},\mathcal{S}\setminus\left\{s\right\}}.
\end{equation}
Thus, with a single multicast message the server can deliver $t+1$ sub-files, and achieve a {\em multicasting gain} of $t+1$.

\section{Cross-level Coded Delivery
(CLCD) SCHEME}\label{s:CLCD}
\subsection{An introductory overview}
In the proposed CLCD scheme, the file library is divided into three groups according to their popularity: the most popular $N_{h}$ files are called the {\em high-level} files, the less popular $N_l$ files are called the {\em low-level} files, and the remaining least popular $N_r=N-(N_h+N_l)$ files are called  the {\em zero-level} files. All the files in the same group are cached at the same level. Hence, the scheme can be represented by five parameters $t_{h},t_{l},N_{h},N_{l},N_{r}$, where $t_{h}$ and $t_{l}$ represent the caching levels of the low-level and high-level files, respectively, whereas  the zero-level files are not cached at all. Hence, we use $CL(t_h,t_l,N_h,N_l,N_r)$ to denote the overall caching strategy.\\
\indent For the given parameters $t_{h},t_{l},N_{h},N_{l},N_{r}$,  the required normalized cache-memory size $M(t_{h},t_{l},N_{h},N_{l},N_{r})$ is given by
\begin{equation}
M\left( t_{h}, t_{l}, N_{h}, N_{l},N_{r}\right) \defeq  \frac{t_{h}N_{h}+t_{l}N_{l}}{K}\label{cache_cap},
\end{equation}
since caching a file at level $t$ means dividing it into ${K \choose t}$ sub-files and letting each user cache  ${K-1 \choose t-1}$ sub-files; that is, each user caches $t/K$ portion of each file.\\
\indent The key contribution of the CLCD scheme is in the delivery phase. When the user demands are revealed, they are clustered into three groups,  denoted by $\mathcal{K}^{h}$, $\mathcal{K}^{l}$ and $\mathcal{K}^{r}$, which correspond to the set of users that request high-level, low-level, and  zero-level files, respectively. A naive approach would be to serve the users in each set $\mathcal{K}^{a}$, $a\in\left\{h,l,r\right\}$, separately as in Algorithm \ref{alg:naive}.
\begin{algorithm}[t]
    \SetKwInOut{Input}{Input}
    \SetKwInOut{Output}{Output}
    \Input{$\mathcal{K}^{h}$, $\mathcal{K}^{l}$}
		\For{$a\in\left\{h,l\right\}$}{
		    \For{$\mathcal{S}\subseteq:\vert\mathcal{S}\vert=t_{a}$}{
		        \If{$\mathcal{S}\cap\mathcal{K}^{a}\neq\emptyset$}{
		        Multicast $\bigoplus_{s\in \mathcal{K}^{a}}W_{d_{s},\mathcal{S}\setminus\left\{s\right\}}$
		        }
		        
		    }
		}
Serve users in $\mathcal{K}^{r}$ with unicast messages.
\caption{Naive coded delivery for non-uniform content placement}
\label{alg:naive}
\end{algorithm}
The main limitation of this approach is that, although multicasting gain of $t_{a}+1$ is targeted at line 4 of Algorithm \ref{alg:naive}, the achieved multicasting gain is limited to  $\vert\mathcal{S}\cap\mathcal{K}^{a}\vert$. The main objective of the CLCD strategy is to design a proper way of forming multicast messages such that  users from two different groups can be targeted as well  in order to reduce the communication load. To this end, CLCD addresses two main challenges; first, how to form the set of target users for multicating, second, how to align sub-file sizes. The latter challenge stems from the mismatch between the sizes of high-level and low-level sub-files. This mismatch can be resolved either at the placement or at the delivery phase. That is, the size of each multicast message can be fixed, or can be adjusted during the delivery phase according to the targeted users. This will be further clarified later in the paper. To simplify the presentation and gradually introduce the key ideas behind the CLCD, in the following subsections we will start introducing special cases of the general CLCD scheme, and gradually extend the analysis to the more general version.

\subsection{$\mathrm{CL(t_h=2,t_l=1,N_h, N_l, N_{r} =0)}$ }\label{ss:CL_2_1}
In this section, we focus on  a particular case of  CLCD scheme, $CL(2,1,N_{h},N_{l},0)$, that is high-level files are cached at level $t_h=2$ and low-level files are cached at level $t_l=1$, $N_h = N-N_l$, that is, all the files are cached at either at level 1 or level 2. We will later extend our analysis to the more general CLCD schemes.
\subsubsection{Placement phase}\label{ss:placement}
We first want to remark that the placement strategy employed here is not specific to $t_h=2$ and can be utilized for more general schemes as well, as long as $t_l=1$. The placement of the high-level files is identical to the one in \cite{CD.F1} such that each file is divided into ${K \choose t_h}$ sub-files and the high-level files are cached at level $t_h$; that is, each user caches ${K-1 \choose t_h-1}$ sub-files of each high-level file. 

The key difference appears in the placement of the low-level files; instead of dividing files into ${K \choose 1}$ sub-files, we use the same sub-packetization scheme used for the high-level files above and group them into $K$ disjoint and equal-size subsets so that each user caches the sub-files of a different subset. Equivalently, each sub-file is cached by only a single user, and each user caches only ${K-1 \choose t_{h}-1}/t_{h}$ sub-files of each low-level file exclusively. We assume here that ${K \choose t_{h}}$ is divisible by $K$, which holds for any $t_{h}$ when $K$ is a prime number. We emphasize that this assumption is not required for CLCD in general; however, under this assumption all the files are divided into equal number of sub-files, thus the alignment problem can be resolved at the placement phase as we explain next. \\
\indent Although the presented placement scheme is general, in the case of $t_h=2$, we can provide a closed form expression for the achievable delivery rate as a function of the requested files from each caching level, which is denoted by  $k^{a}\defeq \vert \mathcal{K}^{a}\vert, a\in\left\{l,h,r\right\}$, and the corresponding delivery strategy can be obtained in  polynomial time. This is particularly important, since it implies a low complexity coded delivery design with a certain achievable rate guarantee in the low cache-memory regime. Next, we provide an example to better illustrate the proposed placement strategy.

\begin{table}
{\footnotesize
  \centering
  \begin{tabular}{|M{1cm}|M{0.65cm}|M{0.65cm}|M{0.65cm}|M{0.65cm}|M{0.65cm}|M{0.65cm}|M{0.65cm}|}
    \hline
      User & 1  & 2  & 3 & 4 & 5 & 6 & 7 \\ \hline
  index of cached sub-files & {\color{red}$\left\{1,2\right\}$ $\left\{1,3\right\}$ $\left\{1,4\right\}$} {\color{blue}$\left\{1,5\right\}$ $\left\{1,6\right\}$ $\left\{1,7\right\}$} & {\color{blue}$\left\{1,2\right\}$} {\color{red}$\left\{2,3\right\}$ $\left\{2,4\right\}$ $\left\{2,5\right\}$} {\color{blue}$\left\{2,6\right\}$ $\left\{2,7\right\}$} & {\color{blue}$\left\{1,3\right\}$ $\left\{2,3\right\}$} {\color{red}$\left\{3,4\right\}$ $\left\{3,5\right\}$ $\left\{3,6\right\}$} {\color{blue}$\left\{3,7\right\}$} & {\color{blue}$\left\{1,4\right\}$ $\left\{2,4\right\}$ $\left\{3,4\right\}$} {\color{red}$\left\{4,5\right\}$ $\left\{4,6\right\}$ $\left\{4,7\right\}$} & {\color{red}$\left\{1,5\right\}$} {\color{blue}$\left\{2,5\right\}$ $\left\{3,5\right\}$ $\left\{4,5\right\}$} {\color{red}$\left\{5,6\right\}$ $\left\{5,7\right\}$} & {\color{red}$\left\{1,6\right\}$ $\left\{2,6\right\}$} {\color{blue}$\left\{3,6\right\}$ $\left\{4,6\right\}$ $\left\{5,6\right\}$} {\color{red}$\left\{6,7\right\}$} & {\color{red}$\left\{1,7\right\}$ $\left\{2,7\right\}$ $\left\{3,7\right\}$} {\color{blue}$\left\{4,7\right\}$ $\left\{5,7\right\}$ $\left\{6,7\right\}$} \\ \hline 
\end{tabular}
  \newline\newline
  \caption{Sub-file placement: for each user the sub-files in red are cached for all the files, whereas the sub-files in blue are cached for only high-level files.}
	\label{subfile}
	}
\end{table}

\textbf{Example 1.} Consider a network of $K=7$ users and a library of $N=7$ files  with decreasing popularity. To simplify the notation, we will denote the files as  $W_{1}=A$, $W_{2}=B$, $W_{3}=C$, $W_{4}=D$, $W_{5}=E$, $W_{6}=F$, $W_{7}=G$. We set the normalized  cache size as $M=12/7$. For the given cache size, we have $N_{h}=5$; that is, the most popular files, $\left\{A,B,C,D,E\right\}$, are treated as  high-level files, whereas,  $G$ and $F$ as low level files. Each file is divided into ${K \choose 2}=21$ sub-files, and each user stores $K-1=6$ sub-files of each high-level file, and $\frac{K-1}{2}=3$ sub-files of each low-level file. User cache contents after the placement phase is illustrated in Table \ref{subfile}; for each user the sub-files in red are cached for all the files, whereas the sub-files in blue are cached for only the high-level files.
\begin{remark}
 In Example 1, we use a {\em systematic placement} strategy for the low-level files; such that, for each user, the index set of the cached sub-files of a low-level file is a subset of the index set of the cached sub-files of a high level file. Although, this particular  placement strategy may not affect the delivery rate performance, it can be beneficial in practice if the library and/or the file popularity dynamically varies over time and the cached content should be updated accordingly \cite{DCU}. If a high-level file becomes low-level, then the user simply removes the sub-files colored with blue, and similarly, if a low-level file becomes high-level, then the user only downloads the sub-files colored with blue from the server.  This systematic placement strategy can be summarized in the following way. Let $I_{i}=\left\{\mathcal{S}:  \vert\mathcal{S}\vert = 2,  \mathcal{S}\subseteq \mathcal{K}, i\in\mathcal{S} \right\}$ be the set of index sets of sub-files that contain $i$. Then, when a file $W_{k}$ is cached at level 2, user $i$ caches all the sub-files $W_{k,\mathcal{S}}$, $\mathcal{S} \in I_{i}$. Further, let each $I_{i}$, for $i\in[K]$, be ordered as illustrated in Table \ref{subfile}, where the $i$th column includes the elements of $I_{i}$ within an order. Once we have ordered sets for each user, then, starting from the first user, each user selects the first $\frac{K-1}{2}$ index set, that are not chosen, in order to cache corresponding low-level files as illustrated in Table \ref{subfile} for $K=7$.
\end{remark}

\subsubsection{Delivery phase}\label{ss:delivery}
\begin{table*}{\footnotesize
     \begin{center}
    \begin{tabular}{ | M{3cm} | M{3cm} | M{3cm} | M{3cm} |}
    \hline
    High-level users & Multicast messages & Low-level users &  Multicast messages \\ \hline
        {1 2 3} & {\color{red} $A_{23} \oplus B_{13} \oplus C_{12}$} & {6 7} & {\color{red}$G_{16} \oplus F_{17}$} \\ \hline 
		{1 2 4} & {\color{red}$A_{24} \oplus B_{14} \oplus D_{12}$} & {6 7} & {\color{red}$G_{26} \oplus F_{27}$}\\ \hline
		{1 2 5} & {\color{red}$A_{25} \oplus B_{15} \oplus E_{12}$} & {6 7} & {\color{red}$G_{67} \oplus F_{37}$}\\ \hline
		{1 3 4} & {\color{red}$A_{34} \oplus C_{14} \oplus D_{13}$} & {6} &  {\color{blue}$F_{12}$}\\ \hline
		{1 3 5} & {\color{red}$A_{35} \oplus C_{13} \oplus E_{13}$} & {6} &  {\color{blue}$F_{13}$} \\ \hline
		{1 4 5} & {\color{red}$A_{45} \oplus D_{15} \oplus E_{14}$} & {6} &  {\color{blue}$F_{14}$} \\ \hline
		{2 3 4} & {\color{red}$B_{34} \oplus C_{24} \oplus D_{23}$} & {6} &  {\color{blue}$F_{23}$} \\ \hline
		{2 3 5} & {\color{red}$B_{35} \oplus C_{25} \oplus E_{23}$} & {6} &  {\color{blue}$F_{24}$} \\ \hline
		{2 4 5} & {\color{red}$B_{45} \oplus D_{25} \oplus E_{24}$} & {6} &  {\color{blue}$F_{25}$} \\ \hline
		{3 4 5} & {\color{red}$C_{45} \oplus D_{35} \oplus E_{34}$} & {6} & {\color{blue}$F_{34}$} \\ \hline 
        {1 2 } & {\color{blue} $A_{26} \oplus B_{16}$} & {6} &  {\color{blue}$F_{35}$} \\ \hline
        {1 2 } & {\color{green} $A_{27} \oplus B_{17}$} & {6} & {\color{blue}$F_{36}$} \\ \hline 
        {1 3 } & {\color{blue}$A_{36} \oplus C_{16}$} & {6} &  {\color{blue}$F_{45}$} \\ \hline
        {1 3 } & {\color{blue}$A_{37} \oplus C_{17}$} & {6} &  {\color{blue}$F_{46}$} \\ \hline
        {1 4 } & {\color{blue}$A_{46} \oplus D_{16}$} & {6} &   {\color{blue}$F_{47}$} \\ \hline
        {1 4 } & {\color{green}$A_{47} \oplus D_{17}$} & {6} &  {\color{blue}$F_{15}$} \\ \hline
        {1 5 } & {\color{blue}$A_{56} \oplus E_{16}$} & {6} &  {\color{blue}$F_{56}$} \\ \hline
        {1 5 } & {\color{blue}$A_{57} \oplus E_{17}$} & {6} &  {\color{blue}$F_{57}$} \\ \hline
        {2 3 } & {\color{blue}$B_{36} \oplus C_{26}$} & {7} &  {\color{blue}$G_{12}$} \\ \hline
        {2 3 } & {\color{blue}$B_{37} \oplus C_{27}$} & {7} &   {\color{blue}$G_{13}$} \\ \hline
        {2 4 } & {\color{blue}$B_{46} \oplus D_{26}$} & {7} &   {\color{blue}$G_{14}$} \\ \hline
        {2 4 } & {\color{blue}$B_{47} \oplus D_{27}$} & {7} &  {\color{blue}$G_{23}$} \\ \hline
        {2 5 } & {\color{green} $B_{56} \oplus E_{26}$} & {7} &  {\color{blue}$G_{24}$} \\ \hline
        {2 5 } & {\color{green} $B_{57} \oplus E_{27}$} & {7} &  {\color{blue}$G_{25}$} \\ \hline 
        {3 4 } & {\color{green}$C_{46} \oplus D_{36}$} & {7} &  {\color{blue}$G_{34}$} \\ \hline
        {3 4 } & {\color{green}$C_{47} \oplus D_{37}$} & {7} &  {\color{blue}$G_{35}$} \\ \hline
        {3 5 } & {\color{blue}$C_{56} \oplus E_{36}$} & {7} &  {\color{blue}$G_{36}$} \\ \hline
        {3 5 } & {\color{green}$C_{57} \oplus E_{37}$} & {7} &  {\color{blue}$G_{45}$} \\ \hline
        {4 5 } & {\color{blue}$D_{56} \oplus E_{46}$} & {7} &  {\color{blue}$G_{46}$} \\ \hline
        {4 5 } & {\color{blue}$D_{57} \oplus E_{47}$} & {7} &  {\color{blue}$G_{47}$} \\ \hline
        {1 } & {\color{blue}$A_{67}$} & {7} &  {\color{blue}$G_{15}$} \\ \hline
        {2 } & {\color{blue}$B_{67}$} & {7} & {\color{blue}$G_{56}$} \\ \hline
        {3 } & {\color{blue}$C_{67}$} & {7} & {\color{blue}$G_{57}$} \\ \hline
        {4 } & {\color{blue}$D_{67}$} & - & -\\ \hline
        {5 } & {\color{blue}$E_{67}$} & - & -\\ \hline
		\end{tabular}
\end{center}\caption{Multicast messages constructed according to the conventional coded  delivery scheme for Example 1.}
\label{Conv_delivery}		          
		       }
\end{table*}
Delivery phase of the proposed scheme consists of  four steps. We will continue to use Example 1 to explain the key ideas of the delivery scheme\\
\textbf{Example 1 (continued).} 
Assume that user $k$ requests file $W_{k}$, $k\in[K]$, so that  $\mathcal{K}^{h}=\left\{1,2,3,4,5\right\}$ and $\mathcal{K}^{l}=\left\{6,7\right\}$. Before explaining the delivery steps in detail, we want to highlight the main idea behind the CLCD scheme, which has been inspired  by the {\em bit borrowing} approach introduced in \cite{CD.ND7}. For the given example, if the conventional coded delivery scheme is used for delivering the missing sub-files of the high-level and low-level users separately, then the server should send all the messages given in Table \ref{Conv_delivery}. Now, consider the following three messages  in Table \ref{Conv_delivery}: $A_{26} \oplus B_{16}$, which targets  high-level users 1 and 2, and $F_{12}$ and $F_{23}$, which are sent as unicast messages to low-level user 6. In the cross-level design, the server decomposes the message $A_{26} \oplus B_{16}$ into sub-files $A_{26}$ and $B_{16}$, and instead, pairs them with sub-files 
$F_{12}$ and  $F_{23}$ to construct cross-level messages $A_{26}\oplus F_{12}$, which targets high-level user 1 and low-level user 6, and $B_{16}\oplus F_{23}$, which targets high-level user 2 and low-level user 6. Hence, the server multicasts two messages instead of three. We call this process as {\em multicast message decomposition}, since the cross-level messages are constructed by decomposing a message constructed according to the conventional coded delivery scheme.

The key challenge in CLCD is deciding which messages to decompose, and how to pair the decomposed high-level sub-files with the low-level sub-files. Now, if we go back to Example 1, all the messages in {\color {blue}blue} in Table \ref{Conv_delivery} will be used for cross-level delivery and the total number of broadcast messages will be reduced to $51$ from $68$ (24\% reduction). In the delivery phase, the server first multicasts the messages  in {\color {red}red} in Table \ref{Conv_delivery}, then the messages in {\color {blue}blue} are used to construct cross-level messages, and finally the remaining messages in {\color {green}green} are sent as in the conventional coded delivery scheme. Now, we will further explain the four steps of the delivery phase:\\  
\textbf{Step 1) Intra-high-level delivery:}
The first step of the delivery phase is identical to that in (\ref{coddel}) for $t=2$. The only difference is that, now we consider only the users in $\mathcal{K}^{h}$, instead of $\left[K\right]$. This corresponds to the {\color {red}red} messages in the second column of Table \ref{Conv_delivery}.
\\
\textbf{Step 2) Intra-low-level delivery:}
The second step also follows (\ref{coddel}) with $t=1$, targeting low-level users in $\mathcal{K}^{l}$. These are the {\color {red}red} messages in the last column of Table \ref{Conv_delivery}.\\ 
\textbf{Step 3) Cross-level delivery:}
This step is the main novelty of the CLCD scheme. First, note that each high-level user has $(K-1)/2$ sub-files in its cache that are requested by a low-level user. For instance, in Example 1, user 1 has sub-files $\left\{G_{12},G_{13},G_{14}\right\}$ that are requested by user 7. For $i\in \mathcal{K}^{h}$ and $j\in \mathcal{K}^{l}$, let $\mathcal{H}_{i,j}$ denote the set of sub-files stored at high-level user $i$ and requested by low-level user $j$, e.g., $\mathcal{H}_{1,7} \defeq \left\{G_{12},G_{13},G_{14}\right\}$ in Example 1. We note that these sub-files are sent as unicast messages in the conventional coded delivery scheme. Similarly, for $i\in \mathcal{K}^{h}$ and $j\in \mathcal{K}^{l}$, let $\mathcal{L}_{i,j}$ be the set of ${K-2 \choose 1}=K-2$ sub-files stored by low-level user $j$ that are requested by high-level user $i$, e.g., $\mathcal{L}_{1,7} \defeq \left\{A_{27},A_{37},A_{47},A_{57},A_{67}\right\}$ in Example 1. One can observe that a cross-level message, targeting high-level user $i$ and low-level user $j$, can be constructed by taking one sub-file from each of $\mathcal{H}_{i,j}$ and $\mathcal{L}_{i,j}$, and bit-wise XORing them. For the given example, any three sub-files can be chosen from set 
$\mathcal{L}_{1,7}$ and  paired with any of the sub-files in  $\mathcal{H}_{1,7}$ to construct a cross-level message. To generalize, if there is a set of sub-files $\mathcal{F}_{i,j}\subseteq \mathcal{L}_{i,j}$ such that $\vert\mathcal{F}_{i,j}\vert=\vert \mathcal{H}_{i,j} \vert$, then we can easily construct cross-level messages that target high-level user $i$ and low-level user $j$,  using any one-to-one mapping between these two sets.  However, we remark that, if  sub-file $A_{37}$ is used to construct a cross-level message, i.e., $A_{37}\in \mathcal{F}_{1,7}$, then the multicast message $A_{37}\oplus C_{17}$ should also be decomposed, and the sub-file $C_{17}$ should also be paired with a low-level file, i.e., $C_{17} \in \mathcal{F}_{3,7}$.\\
\indent Hence, the main challenge in cross-level delivery is the construction of sets $\mathcal{F}_{i,j}$ in a joint manner. Before the construction process of these sets, we will define two more sets that will be useful. We note that, in the given example, sub-files $\left\{ A_{27},A_{37},A_{47}, A_{57}\right\}$ are also cached by a high-level user, but sub-file $A_{67}$ is cached by only low-level users. For $i\in \mathcal{K}^{h}$ and $j\in \mathcal{K}^{l}$, we introduce the set $\Omega_{i,j}\subseteq \mathcal{L}_{i,j}$ of the sub-files that are requested by high-level user $i$, and cached by low-level user $j$ as well as by a high-level user, e.g., $\Omega_{1,7} \defeq \left\{A_{27},A_{37},A_{47},A_{57}\right\}$. Furthermore, we introduce the set $\Lambda_{i}$, $i\in \mathcal{K}^{h}$, of the sub-files that are requested by  high-level user $i$ but cached by only low-level users, i.e., $\Lambda_{i}\defeq\cup_{j\in\mathcal{K}^{l}}\left(\mathcal{L}_{i,j}\setminus\Omega_{i,j}\right)$. We have, in Example 1, $\Lambda_{1}=\left\{A_{67}\right\}$.\\ 
\indent In the third step of proposed CLCD scheme, our aim is to deliver all the remaining sub-files requested by the low-level users and the sub-files requested by the high-level users that are cached by only low-level users, via multicast messages, each destined for one high-level and one low-level user. More formally, we want low-level user $j$,  $j\in\mathcal{K}^{l}$, to recover all the sub-files in $\cup_{i\in \mathcal{K}^{h}} \mathcal{H}_{i,j}$, and we want high-level user $i$, $i\in\mathcal{K}^{h}$, to recover all the sub-files in $\Lambda_{i}$. To this end, sets $\mathcal{F}_{i,j}$ must satisfy the following properties:
\begin{align}
&\Lambda_{i}\subseteq \cup_{j} \mathcal{F}_{i,j}, \text{  } \forall  i\in\mathcal{K}^{h}, \label{complete}\\
& \mathcal{F}_{i,j}\cap \mathcal{F}_{i,k}= \emptyset, \text{  } \forall i \in\mathcal{K}^{h} \text{ and } \forall j,k\in\mathcal{K}^{l} \label{exclusive},
\end{align}
where (\ref{complete}) ensures that each high-level user collects its missing sub-files that are  available only in the caches of the low-level users and (\ref{exclusive}) guarantees that  high-level users do not receive the same sub-file multiple times.\\
\indent Next, we show how to construct the sets $\mathcal{F}_{i,j}$, $i\in\mathcal{K}^{h}$, $j\in\mathcal{K}^{l}$. In order to ensure (\ref{complete}) and (\ref{exclusive}), $\Lambda_{i}$ is partitioned into subsets $\left\{\Lambda_{i,j}\right\}_{j\in K^{l}}$ with approximately uniform cardinality, i.e., $\vert\Lambda_{i,k}\vert-\vert\Lambda_{i,j}\vert\leq 1$, $\forall j,k\in \mathcal{K}^{l}$, $j\neq k$, and such that $\Lambda_{i,j}\subseteq \mathcal{L}_{i,j}$ holds for all $i\in\mathcal{K}^{h}$ and $j\in\mathcal{K}^{l}$. Further details on  approximately uniform partitioning are provided in Appendix \ref{approx_part}. We note that, if $\Lambda_{i,j} \subseteq \mathcal{F}_{i,j}$, then (\ref{complete}) holds. We also assume that the same partitioning is applied to all $\Lambda_{i}$'s. Partitions of $\Lambda_{i}$'s for Example 1 are illustrated in Table \ref{sets}.\\ 
\indent For  given $\Omega_{i,j}$ and $\Lambda_{i,j}$, $\mathcal{F}_{i,j}$ can be constructed as follows:
$\mathcal{F}_{i,j}=\Lambda_{i,j} \cup \left\{\Omega_{i,j} \setminus \Delta_{i,j}\right\}$,
for some $\Delta_{i,j}\subseteq\Omega_{i,j}$. Then, from the construction, one can  easily verify that (\ref{exclusive}) holds. We note that, in this construction, $\Delta_{i,j}$ simply denotes the sub-files in $\mathcal{L}_{i,j}$ that are not used in a cross-level message. In Example 1, this corresponds to the green multicast  messages in Table \ref{Conv_delivery}. In particular, if multicast message $A_{27} \oplus B_{17}$ is not decomposed for cross-level delivery, then  $A_{27}\in \Delta_{17}$ and $B_{17}\in\Delta_{27}$. Hence, the ultimate aim is to find a proper way of constructing sets $\Delta_{ij}$, which is equivalent to deciding which multicast messages to be decomposed.\\
\indent In order to ensure the initial requirement of $\vert\mathcal{F}_{i,j}\vert=\vert \mathcal{H}_{i,j} \vert$, we need to show that it is possible to construct $\Delta_{i,j}$, $i\in\mathcal{K}^{h}$, $j\in\mathcal{K}^{l}$, which satisfy the following equality
\begin{equation}\label{sufconddelt}
\vert \mathcal{H}_{i,j} \vert= \vert \Lambda_{i,j}\vert+ \vert \Omega_{i,j} \vert- \vert \Delta_{i,j}\vert.
\end{equation}
We note that, if the following inequality holds  
\begin{equation}
\vert \Lambda_{i,j}\vert \leq \vert \mathcal{H}_{i,j} \vert \leq \vert \Lambda_{i,j}\vert+\vert \Omega_{i,j} \vert,
\end{equation}
 $\Delta_{i,j}$ satisfying (\ref{sufconddelt}) can always be found.\\ 
\indent From the construction, we know that $\vert \mathcal{H}_{i,j} \vert = \frac{K-1}{2}$; however, $\vert \Omega_{i,j} \vert$ and  $\vert \Lambda_{i,j}\vert$ depend on the realization of the user demands, i.e.,  $\vert \Omega_{i,j} \vert= k^{h}-1$ and $\vert \Lambda_{i,j}\vert= \left\lceil \frac{k^{l}-1}{2}\right\rceil$ or $\vert \Lambda_{i,j}\vert= \left\lfloor \frac{k^{l}-1}{2}\right\rfloor$ due to the approximately uniform partitioning. Accordingly, 
\begin{align}
\vert \Omega_{i,j} \vert+ \vert \Lambda_{i,j}\vert& =  \left\lfloor \frac{K-1}{2} + \frac{k^{h}-2}{2} \right\rfloor, \text{ or } \\
\vert \Omega_{i,j} \vert+ \vert \Lambda_{i,j}\vert& =  \left\lceil  \frac{K-1}{2} + \frac{k^{h}-2}{2} \right\rceil.
\end{align}
One can observe that, when $k^{h}\geq 2$, $\vert \Omega_{i,j} \vert+ \vert \Lambda_{i,j} \vert\geq \vert\mathcal{H}_{i,j}\vert=\frac{K-1}{2}$ in both cases. If $k^{h}=1$, the high-level file is considered as a low-level file in the delivery phase, and the achievable rate becomes $(K-1)/2$. In the remainder of this section, we assume $k^{h}\geq 2$. One can also observe that $\vert \Lambda_{i,j}\vert \leq \vert \mathcal{H}_{i,j}\vert$, since $k^{l}\leq K$. Let $n_{i,j}$ be the cardinality of the set $\Delta_{i,j}$ that satisfies (\ref{sufconddelt}), i.e., $n_{i,j}\defeq \vert \Omega_{i,j} \vert + \vert \Lambda_{i,j}\vert-\vert \mathcal{H}_{i,j} \vert$. We can consider any subset of $\Omega_{i,j}$ with cardinality $n_{i,j}$ as $\Delta_{i,j}$ to construct $\mathcal{F}_{i,j}$.
\begin{table*}
{\footnotesize
    \centering
    \begin{tabular}{| M{1cm} | M{2.5cm} | M{1cm} |M{1.5cm} |M{2cm}| M{2cm} | M{4cm} |}
    \hline
(i,j) &	$\Omega_{i,j}$ &	$\Lambda_{i,j}$ & $\Delta_{i,j}$ & $\mathcal{F}_{i,j}$ & $\mathcal{H}_{i,j}$ & Cross-level messages \\ \hline
(1,6) &  $\left\{{\color{blue}A_{26},A_{36}},{\color{red} A_{46}},{\color{blue} A_{56}}\right\}$ & $\emptyset$ & $\left\{{\color{red}A_{46}}\right\}$ & $\left\{{\color{blue}A_{26},A_{36},A_{56}}\right\}$  & $\left\{F_{12},F_{13},F_{14}\right\}$ & $A_{26} \oplus F_{12}$, $A_{36} \oplus F_{13}$, $A_{56} \oplus F_{14}$ \\ \hline
(1,7) &	 $\left\{{\color{red}A_{27}},{\color{blue}A_{37}},{\color{red} A_{47}},{\color{blue} A_{57}}\right\}$ &	$\left\{{\color{blue}A_{67}}\right\}$ & $\left\{{\color{red}A_{27},A_{47}}\right\}$ &$\left\{{\color{blue}A_{67},A_{37},A_{57}}\right\}$  & $\left\{G_{12},G_{13},G_{14}\right\}$ & $A_{67} \oplus G_{12}$, $A_{37} \oplus G_{13}$, $A_{57} \oplus G_{14}$ \\ \hline 
(2,6) &		$\left\{{\color{blue}B_{16},B_{36},B_{46}},{\color{red} B_{56}}\right\}$ &$\emptyset$&	$\left\{{\color{red}B_{56}}\right\}$ & $\left\{{\color{blue}B_{16},B_{36},B_{46}}\right\}$  & $\left\{F_{23},F_{24},F_{25}\right\}$ & $B_{16} \oplus F_{23}$, $B_{36} \oplus F_{24}$, $B_{46} \oplus F_{25}$ \\ \hline 
(2,7) &			$\left\{{\color{red}B_{17}},{\color{blue}B_{37},B_{47}},{\color{red} B_{57}}\right\}$ & $\left\{{\color{blue}B_{67}}\right\}$ & $\left\{{\color{red}B_{17},B_{57}}\right\}$ & $\left\{{\color{blue}B_{67},B_{37},B_{47}}\right\}$  & $\left\{G_{23},G_{24},G_{25}\right\}$ & $B_{67} \oplus G_{23}$, $B_{37} \oplus G_{24}$, $B_{47} \oplus G_{25}$ \\ \hline
(3,6) &		$\left\{{\color{blue}C_{16},C_{26}},{\color{red}C_{46}},{\color{blue} C_{56}}\right\}$ &$\emptyset$ & $\left\{{\color{red}C_{46}}\right\}$  & $\left\{{\color{blue}C_{16},C_{26},C_{56}}\right\}$  & $\left\{F_{34},F_{35},F_{36}\right\}$ & $C_{16} \oplus F_{34}$, $C_{26} \oplus F_{35}$, $C_{56} \oplus F_{36}$ \\ \hline  
(3,7) &	$\left\{{\color{blue}C_{17},C_{27}},{\color{red}C_{47}, C_{57}}\right\}$ & $\left\{{\color{blue}C_{67}}\right\}$ & $\left\{{\color{red}C_{47},C_{57}}\right\}$ & $\left\{{\color{blue}C_{67},C_{17},C_{27}}\right\}$  & $\left\{G_{34},G_{35},G_{36}\right\}$ & $C_{67} \oplus G_{34}$, $C_{17} \oplus G_{35}$, $C_{27} \oplus G_{36}$ \\ \hline
 (4,6) & $\left\{{\color{blue}D_{16},D_{26}},{\color{red}D_{36}},{\color{blue} D_{56}}\right\}$ & $\emptyset$ & $\left\{{\color{red}D_{36}}\right\}$ & $\left\{{\color{blue}D_{16},D_{26},D_{56}}\right\}$  & $\left\{F_{45},F_{46},F_{47}\right\}$ & $D_{16} \oplus F_{45}$, $D_{26} \oplus F_{46}$, $D_{56} \oplus F_{47}$ \\ \hline
	(4,7) & $\left\{{\color{red}D_{17}},{\color{blue}D_{27}},{\color{red}D_{37}},{\color{blue} D_{57}}\right\}$	& $\left\{{\color{blue}D_{67}}\right\}$ & $\left\{{\color{red}D_{17},D_{37}}\right\}$ & $\left\{{\color{blue}D_{67},D_{27},D_{57}}\right\}$  & $\left\{G_{45},G_{46},G_{47}\right\}$ & $D_{67} \oplus G_{45}$, $D_{27} \oplus G_{46}$,$D_{57} \oplus G_{47}$ \\ \hline 
	(5,6) & $\left\{{\color{blue}E_{16}},{\color{red}E_{26}},{\color{blue}E_{36}, E_{46}}\right\}$		&$\emptyset$ & $\left\{{\color{red}E_{26}}\right\}$ & $\left\{{\color{blue}E_{16},E_{36},E_{46}}\right\}$  & $\left\{F_{15},F_{56},F_{57}\right\}$  & $E_{16} \oplus F_{15}$, $E_{36} \oplus F_{56}$, $E_{46} \oplus F_{57}$ \\ \hline
	(5,7) &	$\left\{{\color{blue}E_{17}},{\color{red}E_{27},E_{37}}, {\color{blue}E_{47}}\right\}$	&$\left\{{\color{blue}E_{67}}\right\}$ & $\left\{{\color{red}E_{27},E_{37}}\right\}$ & $\left\{{\color{blue}E_{67},E_{17},E_{47}}\right\}$  & $\left\{G_{15},G_{56},G_{57}\right\}$ & $E_{67} \oplus G_{15}$, $E_{17} \oplus G_{56}$, $E_{47} \oplus G_{57}$ \\ \hline
    \end{tabular}
		\caption{Sets $\Omega_{i,j}$, 	$\Lambda_{i,j}$,  $\Delta_{i,j}$,  $\mathcal{F}_{i,j}$,  $\mathcal{H}_{i,j}$, and the cross-level messages formed in Step 3 of the delivery phase for Example 1.}
		\label{sets}		
}		        
\end{table*}
 Eventually, the overall problem can be treated as choosing the multicast messages that will be decomposed, colored with blue in Table \ref{Conv_delivery}, so that the sub-files that are not used for cross-level delivery satisfy the constraint $\vert \Delta_{i,j} \vert= n_{i,j}$, $\forall$ $i\in\mathcal{K}^{h},j\in\mathcal{K}^{l}$. Once the decomposed files and $\Delta_{i,j}$ are decided, $\mathcal{F}_{i,j}$ can be constructed easily as illustrated in Table \ref{sets}, which also lists all the cross-level messages.\\
\indent  We remark that when $W_{d_{i},\left\{k,j\right\}}\in \Delta_{i,j}$, for $i,k\in \mathcal{K}^{h}$ and $j\in\mathcal{K}^{l}$, this also requires $W_{d_{k},\left\{i,j\right\}}\in \Delta_{k,j}$. Hence, for a particular $j\in\mathcal{K}^{l}$ with $n_{i,j}=n$, for all $i\in\mathcal{K}^{h}$, construction of the sets $\Delta_{i,j}$ is equivalent to finding $k^{h}n/2$ number of $(i,k)$ pairs such that each $i \in \mathcal{K}^{h}$ appears in exactly $n$ of them. This problem can be analyzed by constructing a graph whose vertices are the elements of $\mathcal{K}^{h}$, and each vertex has a degree of $n$ so that each edge of the graph can be considered as a pair. To successfully construct such a graph we utilize the {\em matching} concept used in graph theory. The key graph theoretic definitions and lemmas used for this analysis are introduced in Appendix \ref{appendix_partition}, while the  construction of sets  $\Delta_{i,j}$ and $\mathcal{B}$ is detailed in Appendix \ref{appendix_construc}.

\textbf{Step 4) Intra-high-level delivery with multicasting gain of two:}
In the last step, the server multicasts the remaining messages which are colored with green in Table \ref{Conv_delivery}. In Example 1, this step is finalized with unicasting the sub-file $A_{46}$. Let $\mathcal{B}$ denote the set of multicast messages that are not decomposed to be used for cross-level delivery, and hence, are multicasted in the last step.\\

The required delivery rate for a particular demand vector $\mathbf{d}$ under $\mathrm{CL(2,1,N_{h}, N_{l},0)}$ scheme is given by the following lemma.
\begin{lemma}\label{main_result}
For a cache size of $M=\frac{2N_{h}+N_{l}}{2}$, and demand realization  $\mathbf{d}$, the following delivery rate is achievable by the proposed $\mathrm{CL(2,1,N_{h}, N_{l},0)}$ scheme:
\begin{equation}\label{rate}
    R(\mathbf{d}) =
    \begin{cases*}
      \frac{{k^{h} \choose 3}+ \left\lceil \left(L(\mathbf{d})-3{k^{h} \choose 3}\right)/2\right\rceil}{{K \choose 2}}, & if $k^{h}\geq2$ \\
      \frac{K-1}{2},        & otherwise
    \end{cases*},
  \end{equation} 
where $L(\mathbf{d})$ is the total number of missing sub-files for demand realization $\mathbf{d}$, and is given by
\begin{equation}\label{missfile}
L(\mathbf{d})\defeq \left(K-1\right)\left[\frac{K^{2}}{2}-k^{h}-\frac{k^{l}}{2}\right].
\end{equation}
\end{lemma}

Note that, in the centralized coded delivery scheme of \cite{CD.F1}, when all the files are cached at level $t$, the delivery rate is equal to $\frac{K-t}{t+1}$. We can observe from  (\ref{rate}) that, when $k^{l}=K$, there are only low level users and  $R(\mathbf{d})=\frac{K-1}{2}$. Similarly, when  $k^{h}=K$, there are only high level users and $R(\mathbf{d})=\frac{K-2}{3}$. In general, Lemma \ref{main_result} implies that for any demand realization $\mathbf{d}$, sub-files are served with a minimum multicasting gain of 2. We also remark that the delivery rate depends only on the vector $\mathbf{k} \defeq [k^{h},k^{l}]$; rather than the demand vector; hence, the dependence on $\mathbf{d}$ in (\ref{rate}) and (\ref{missfile}) can be replaced by $\mathbf{k}$, i.e., we use $R(\mathbf{k})$ and $L(\mathbf{k})$, respectively. Note that $k^{h}$ and $k^{l}$ are random variables, and their distribution $P(\mathbf{k}\vert N_h,N_{l})$ depends on $N_h$, $N_l$ and the popularity distribution. Accordingly, the average delivery rate can now be written as follows:
\begin{equation}
\bar{R}\defeq\sum_{\mathbf{k}} R(\mathbf{k}) P(\mathbf{k}\vert N_h,N_l).
\end{equation}

\indent In the next subsection, we extend our analysis to the case in which a certain subset of the files are not cached at all, that is, $N_{h}+N_{l} < N$.
\subsection{$\mathrm{CL(t_h=2,t_l=1, N_h,  N_l, N_r)}$}\label{s:clcd_ext}
 We recall that when $N_{h}+N_{l}=N$, the cache memory requirement for $\mathrm{CL(2,1, N_h, N_l, 0)}$  is given by
\begin{equation}
M\left(2, 1, N_{h}, N_{l}, 0\right) \defeq  \frac{N_{h} + N}{K}\label{cache_cap_2_1}.
\end{equation}
Hence, in this particular case, for a given  cache memory size of $M$, $N_{h}$ and $N_{l}$ can be directly written as a parameter of $N$, $M$, and $K$. However,  by choosing  not to cache $N_{r}$ number of least popular files, it is possible to consider different  values for $N_{h}$ and $N_{l}$ under the constraint
\begin{equation}
  \frac{2N_{h} + N_{l}}{K} \leq M,
\end{equation}
 or, equivalently,
\begin{equation}
  N_{h} - N_{r} \leq KM-N.\label{place210}
\end{equation}
From Equation (\ref{place210}), one can observe that by playing with the parameter $N_{r}$ it is possible to seek a different placement strategy in terms of the caching levels. We will later discuss how the placement strategy can be optimized, but here now we will focus on the delivery  of the $\mathrm{CL(2, 1, N_h,  N_l, N_r)}$ scheme, where we have three set of users,  $\mathcal{K}^{h}$, $\mathcal{K}^{l}$ and $\mathcal{K}^{r}$, and set  $k^{a}\defeq \vert \mathcal{K}^{a}\vert, a\in\left\{l,h,r\right\}$ as before.

\begin{table*}
{\footnotesize
    \centering
    \begin{tabular}{| M{0.7cm} | M{2cm} | M{0.7cm}| M{0.7cm} |M{1.3cm} |M{2cm}| M{2cm} | M{4cm} |}
    \hline
(i,j) &	$\Omega_{i,j}$ & $\Pi_{i,j}$  &	$\Lambda_{i,j}$ & $\Delta_{i,j}$ & $\mathcal{F}_{i,j}$ & $\mathcal{H}_{i,j}$ & Cross-level messages \\ \hline
(1,5) &  $\left\{{\color{red} A_{25}},{\color{blue} A_{35}, A_{45}}\right\}$ & $\left\{{\color{cyan}A_{57}}\right\}$ & $\emptyset$ & $\left\{{\color{red}A_{25}}\right\}$ & $\left\{{\color{cyan}A_{57}},{\color{blue}A_{35},A_{45}}\right\}$  & $\left\{E_{12},E_{13},E_{14}\right\}$ & $A_{57} \oplus E_{12}$, $A_{35} \oplus E_{13}$, $A_{45} \oplus E_{14}$ \\ \hline
(1,6) &	 $\left\{{\color{red}A_{26},A_{36}},{\color{blue} A_{46}} \right\}$ & $\left\{{\color{cyan}A_{67}}\right\}$ &	$\left\{{\color{blue}A_{56}}\right\}$ & $\left\{{\color{red}A_{26},A_{36}}\right\}$ &$\left\{{\color{cyan}A_{67}}, {\color{blue}A_{46},A_{56}}\right\}$  & $\left\{F_{12},F_{13},F_{14}\right\}$ & $A_{67} \oplus F_{12}$, $A_{46} \oplus F_{13}$, $A_{56} \oplus F_{14}$ \\ \hline 
(2,5) &		$\left\{{\color{red}B_{15}}{\color{blue},B_{35},B_{45}}\right\}$ & $\left\{{\color{cyan}B_{57}}\right\}$ & $\emptyset$ &	$\left\{{\color{red}B_{15}}\right\}$ & $\left\{{\color{cyan}B_{57}},{\color{blue}B_{35},B_{45}}\right\}$  & $\left\{E_{23},E_{24},E_{25}\right\}$ & $B_{57} \oplus E_{23}$, $B_{35} \oplus E_{24}$, $B_{45} \oplus E_{25}$ \\ \hline 
(2,6) &	$\left\{{\color{red}B_{16},B_{46}},{\color{blue}B_{36}}\right\}$ & $\left\{{\color{cyan}B_{67}}\right\}$ & $\left\{{\color{blue}B_{56}}\right\}$ & $\left\{{\color{red}B_{16},B_{46}}\right\}$ & $\left\{{\color{cyan}B_{67}},{\color{blue}B_{36},B_{56}}\right\}$  & $\left\{F_{23},F_{24},F_{25}\right\}$ & $B_{67} \oplus F_{23}$, $B_{36} \oplus F_{24}$, $B_{56} \oplus F_{25}$ \\ \hline
(3,5) &		$\left\{{\color{red} C_{45}},{\color{blue} C_{15}, C_{25}}\right\}$ & $\left\{{\color{cyan}C_{57}}\right\}$ & $\emptyset$  & $\left\{{\color{red}C_{45}}\right\}$  & $\left\{{\color{cyan}C_{57}},{\color{blue}C_{15},C_{25}}\right\}$  & $\left\{E_{34},E_{35},E_{36}\right\}$ & $C_{57} \oplus E_{34}$, $C_{15} \oplus E_{35}$, $C_{25} \oplus E_{36}$ \\ \hline  
(3,6) &	$\left\{{\color{red}C_{16},C_{46}},{\color{blue}C_{36}}\right\}$ & $\left\{{\color{cyan}C_{67}}\right\}$ & $\left\{{\color{blue}C_{56}}\right\}$ & $\left\{{\color{red}C_{16},C_{46}}\right\}$ & $\left\{{\color{cyan}C_{67}},{\color{blue}C_{36},C_{56}}\right\}$  & $\left\{F_{34},F_{35},F_{36}\right\}$ & $C_{67} \oplus F_{34}$, $C_{36} \oplus F_{35}$, $C_{56} \oplus F_{36}$ \\ \hline
 (4,5) & $\left\{{\color{red} D_{35}},{\color{blue} D_{25}, D_{15}}\right\}$ &  $\left\{{\color{cyan}D_{57}}\right\}$ & $\emptyset$  & $\left\{{\color{red}D_{35}}\right\}$ & $\left\{{\color{cyan}D_{57}},{\color{blue}D_{15},D_{25}}\right\}$  & $\left\{E_{45},E_{46},E_{47}\right\}$ & $D_{57} \oplus E_{45}$, $D_{15} \oplus E_{46}$, $D_{25} \oplus E_{47}$ \\ \hline
	(4,6) & $\left\{{\color{red}D_{26},D_{36}},{\color{blue}D_{16}}\right\}$	&  $\left\{{\color{cyan}D_{67}}\right\}$ & $\left\{{\color{blue}D_{56}}\right\}$ & $\left\{{\color{red}D_{26},D_{36}}\right\}$ & $\left\{{\color{cyan}D_{67}}, {\color{blue}D_{16},D_{56}}\right\}$  & $\left\{F_{45},F_{46},F_{47}\right\}$ & $D_{67} \oplus F_{45}$, $D_{16} \oplus F_{46}$, $D_{56} \oplus F_{47}$ \\ \hline 
    \end{tabular}
		\caption{Sets $\Omega_{i,j}$, 	$\Lambda_{i,j}$,  $\Delta_{i,j}$,  $\mathcal{F}_{i,j}$,  $\mathcal{H}_{i,j}$, $\Pi_{i,j}$ and the multicasted messages in step 3 of the delivery phase for Example 2.}
		\label{sets2}		
}		        
\end{table*}

\indent We highlight the key modifications compared to the case $N_{r}=0$. First, we introduce a new set $\mathcal{Z}_{i,j}$ of sub-files that are requested by a high-level user $i \in \mathcal{K}^{h}$, and cached by a low-level user $j\in \mathcal{K}^{l}$ and a zero-level user. The fundamental modification  appears in the cross-level delivery step, i.e., in Step 3 of the delivery algorithm,  particularly in the construction of  set $\mathcal{F}_{i,j}$. When $\mathcal{K}^{r}\neq\emptyset$, $\mathcal{F}_{i,j}$ is constructed in the following way   
\begin{equation}
\mathcal{F}_{i,j}=\Lambda_{i,j} \cup \Pi_{i,j} \cup \left\{\Omega_{i,j} \setminus \Delta_{i,j}\right\},
\end{equation}
where $\Pi_{i,j}\subseteq \mathcal{Z}_{i,j}$. Initially, we want all the sub-files in $\mathcal{Z}_{i,j}$ to be included in $\mathcal{F}_{i,j}$, i.e., $\mathcal{Z}_{i,j}\subseteq\mathcal{F}_{i,j}$; however, this is not possible if 
\begin{equation}
\vert \Lambda_{i,j} \vert+\vert \mathcal{Z}_{i,j} \vert >\vert\mathcal{H}_{i,j}\vert.
\end{equation}
In that case, $\sigma_{i,j}$ sub-files must be removed from $\mathcal{Z}_{i,j}$ to obtain $\Pi_{i,j}$, where $\sigma_{i,j}$ is found as 
\begin{align}
\sigma_{i,j}\defeq &  \left( \vert \Lambda_{i,j} \vert+\vert \mathcal{Z}_{i,j} \vert-\vert \mathcal{H}_{i,j}\vert \right)^{+} \nonumber\\
=& \left(\left\lceil \frac{k^{l}-1}{2}\right\rceil + k^{r} - \frac{K-1}{2}
\right)^{+} , \text{or}\nonumber\\
=& \left(
 \left\lfloor \frac{k^{l}-1}{2}\right\rfloor + k^{r} - \frac{K-1}{2}
\right)^{+} ,
\end{align}
where $(x)^{+}$ is $x$ if $x>0$, and $0$ otherwise.

We remark that when $ \sigma_{i,j} \geq 0$, this means that $\vert \Lambda_{i,j} \vert+\vert \Pi_{i,j} \vert= \vert \mathcal{H}_{i,j} \vert$, thus $\Omega_{i,j} = \Delta_{i,j}$ and  $n_{i,j}=\vert\Omega_{i,j}\vert$. In general, $n_{i,j} \vert \Omega_{i,j}\vert-\tilde{n}_{i,j}$, where $\tilde{n}_{i,j}$
\begin{align}
\tilde{n}_{i,j}\defeq & \left(\vert \mathcal{H}_{i,j}\vert-\vert \Lambda_{i,j} \vert-\vert \mathcal{Z}_{i,j} \vert\right)^{+}\nonumber\\
=&\left(\frac{K-1}{2}- \left\lceil \frac{k^{l}-1}{2}\right\rceil - k^{r}\right)^{+}, \text{or}\nonumber\\
=&\left(\frac{K-1}{2}- \left\lfloor \frac{k^{l}-1}{2}\right\rfloor - k^{r} \right)^{+}.
\end{align}
When $n_{i,j}$'s are obtained, the sets $\Delta_{i,j}$ are constructed as in  $\mathrm{CL(2, 1, N_h,  N_l, 0)}$. Overall, the delivery phase of  $\mathrm{CL(2, 1, N_h,  N_l, N_r)}$ consists of five steps. We will use the following example to illustrate these steps.\\
\textbf{Example 2.} In this example, we use the same setup as in Example 1; however, now we assume that files $A,B,C,D$ are high-level files, $E$ and $F$ are low-level files, while  file $G$ is not cached at all. We also assume that users $1,2,3,4,5,6,7$ request  files $A,B,C,D,E,F,G$ respectively. Hence, $\mathcal{K}^{h}=\left\{1,2,3,4\right\}$, $\mathcal{K}^{l}=\left\{5,6\right\}$ and $\mathcal{K}^{r}=\left\{7\right\}$.\\ 
\begin{table}{\footnotesize
    \begin{center}
    \begin{tabular}{ | l | p{3cm} |}
    \hline
    High-level users & Multicast message \\ \hline
    {1 2 3} & $A_{23} \oplus B_{13} \oplus C_{12}$ \\ \hline
		{1 2 4} & $A_{24} \oplus B_{14} \oplus D_{12}$ \\ \hline
		{1 2 } & $A_{27} \oplus B_{17}              $ \\ \hline
		{1 3 4} & $A_{34} \oplus C_{14} \oplus D_{13}$ \\ \hline
		{1 3 } & $A_{37} \oplus C_{17}              $ \\ \hline
		{1 4 } & $A_{47} \oplus D_{17}              $ \\ \hline
		{2 3 4} & $B_{34} \oplus C_{24} \oplus D_{23}$ \\ \hline
		{2 3 } & $B_{37} \oplus C_{27}              $ \\ \hline
		{2 4 } & $B_{47} \oplus D_{27}              $ \\ \hline
		{3 4 } & $C_{47} \oplus D_{37}              $ \\ \hline
     Low-level users & Multicast message \\ \hline
    {5 6} & $E_{16} \oplus F_{15}$ \\ \hline
		{5 6} & $E_{26} \oplus F_{56}$ \\ \hline
		{5 6} & $E_{67} \oplus F_{57}$ \\ \hline	
		\end{tabular}
		\caption{Multicast messages in the first two steps of the delivery phase of  $\mathrm{CL(2, 1, N_h,  N_l, 0)}$ scheme in Example 2.}
		\label{BS2_12}		          
		\end{center}
		       }
\end{table}
\textbf{Step 1) Intra-high-level delivery:}
The first step of the delivery phase is identical to that in (\ref{coddel}) for $t=2$. The only difference is that, now we consider only the users in $\mathcal{K}^{h}\cup\mathcal{K}^{r}$, instead of $\left[K\right]$, and a multicast message is not transmitted for subsets of $\mathcal{K}^{r}$, i.e., $\mathcal{S}\subseteq\mathcal{K}^{r}$.\\
\textbf{Step 2) Intra-low-level delivery:}
The second step also follows (\ref{coddel}) with $t=1$, targeting low-level users in $\mathcal{K}^{l}$. In Example 2, the messages delivered by the server corresponding to the first two steps are listed in  Table  \ref{BS2_12}.\\
\textbf{Step 3) Cross-level delivery:}
We construct sets $\Lambda_{i,j}$ as illustrated in Table \ref{sets2} for Example 2. One can easily observe that $\sigma_{i,j}=0$,   $\forall i\in\mathcal{K}^{h}$ and $\forall j\in\mathcal{K}^{l}$, which implies that $\Pi_{i,j}=\mathcal{Z}_{i,j}$. Then, we evaluate the value of $n_{i,j}$ for all  $i\in\mathcal{K}^{h}$ and $j\in\mathcal{K}^{l}$. Subsequently, we construct the sets $\Delta_{i,j}$ and $\mathcal{F}_{i,j}$ for the evaluated values of $n_{i,j}$. Eventually, sets $\mathcal{F}_{i,j}$ and $\mathcal{H}_{i,j}$ are used to construct the set of multicast messages $\mathcal{B}$ similarly to the case  $\mathcal{K}^{r}=\emptyset$. We again refer the reader to Appendix \ref{appendix_construc} for further details.
Sets $\Delta_{i,j}$, $\mathcal{F}_{i,j}$, and all the multicasted messages in the third step for Example 2 are given in Table \ref{sets2}.\\
\textbf{ Step 4) Intra-high-level delivery with multicasting gain of two:}
In this step, the server multicasts the messages 
\begin{equation}
\begin{aligned}
\mathcal{B}=&\left\{A_{25} \oplus B_{15}, C_{45} \oplus D_{35}, A_{26} \oplus B_{16}, D_{36} \oplus C_{46} \right. \\
&\left. A_{36} \oplus C_{16}, B_{46} \oplus D_{26}\right\},
\end{aligned}
\end{equation}
each of which is destined for two high-level users.\\
\textbf{Step 5) Unicasting:}
The remaining sub-files are sent as unicast messages in the last step. The sub-files that are sent in this step can be categorized under three groups: the first group consists of the  sub-files that are requested by zero-level users, the second group consists of the sub-files that are requested by low-level users and cached by zero-level users, finally the third group is formed by the sub-files in set $\cup_{i\in\mathcal{K}^{h},j\in\mathcal{K}^{l}}(\mathcal{Z}_{i,j}\setminus\Pi_{i,j})$.\\
\indent Let $N_{T}(\mathbf{k})$ be the number of transmitted messages (unicast and multicast) for a given demand realization 
$\mathbf{k}\defeq[k^{h},k^{l},k^{r}]$, i.e., 
\begin{align}
 N_{T}(\mathbf{k})=&\underbrace{{k^{h}+k^{r}\choose 3}-{k^{r}\choose 3}}_{\text{Step 1}}+ \underbrace{\frac{K-1}{2}{k^{l}\choose 2}}_{\text{Step 2}} +\underbrace{\frac{K-1}{2}k^{h}k^{l}}_{\text{Step 3}}\nonumber\\
& +\underbrace{\left\lceil \frac{\sum_{i,j}n_{i,j}}{2}\right\rceil}_{\text{Step 4}}+\underbrace{k^{l}k^{r}\frac{K-1}{2}+\sum_{i,j}\sigma_{i,j}.}_{\text{Step 5}}
\end{align}
Then, the normalized delivery rate for $\mathrm{CL(2, 1, N_h,  N_l, 0)}$ is given by $R(\mathbf{k})=\frac{N_{T}(\mathbf{k})}{{K \choose 2}}+k^{r}$. The average delivery rate can be written as follows

\begin{equation}
\bar{R}(N_h,N_l,N_r)\defeq\sum_{\mathbf{k}} R(\mathbf{k}) P(\mathbf{k}\vert N_{h},N_{l},N_{r}),
\end{equation}
where $\mathbf{k} \defeq [k^{h},k^{l},k^{r}]$ denotes the demand realization. Hence, the placement strategy can be written as an optimization problem;
\begin{small}
\begin{align}
   \text{\bf P1:} \;\;\;\min&
   \begin{aligned}[t]
       \bar{R}(N_h,N_l,N_r) \notag
   \end{aligned} \\
  \text{subject to: }& (\ref{place210})
\end{align}
\end{small}
To solve the optimization problem \text{\bf P1}, we first need to compute the required delivery rate $R(\mathbf{k})$ for each possible demand realization $\mathbf{k}$, which scales with  $\mathcal{O}(K^{2})$. The cache memory constraint in (\ref{place210}) can be rewritten in the following way;
\begin{equation}
N^{max}_{r}\geq N_{r}\geq N^{min}_{r}
\end{equation}
 where $N^{max}_{r}$ and $N^{min}_{r}$ are defined as the number of files that are not cached in order to cache all the remaining files at level $2$ and $1$, respectively. Hence, solving the optimization problem \text{\bf P1} is equivalent to searching for $N_{r}$, that gives the minimum $\bar{R}(N_h,N_l,N_{r})$. Hence, for each value of $N_{r}$, $P(\mathbf{k} \vert N_{h},N_{l},N_{r})$ must be calculated for each possible realization of $\mathbf{k}$, which scales with $\mathcal{O}(K^{2})$. Hence, the  optimization of the placement phase has an overall complexity of $\mathcal{O}(NK^{2}+K^{2})$.


\section{ Extension  to $\mathrm{CL(t_h,t_l=1,N_h, N_l, N_r)}$ } \label{s:clcd_ext}
In this section, we extend our analysis to the more general CLCD scheme where $t_h$ can take any integer value, although we still fix $t_h = 1$. We emphasize here that when $t_h=2$, it is possible to establish a link between the process of constructing multicast messages, particularly forming of the pair of sub-files,  and the concept of {\em matchings} from graph theory. However, this
correspondence does not generalize with $t_h =2$, we particularly focus on a multicasting gain of two, whereas in this section, our aim is provide a more general perspective on CLCD and show how the delivery phase of the proposed scheme can be analyzed as an optimization problem.\\
\indent The placement phase of the $\mathrm{CL(t_h,1,N_h, N_l, N_r)}$ scheme is similar to that of  $\mathrm{CL(2,1,N_h, N_l, N_r)}$. Each file is divided into ${K \choose t_h}$ sub-files, and the high-level files are cached at level $t_h$; that is each user caches ${K-1 \choose t_h-1}$ sub-files of each high-level file. On the other hand, for the 
low-level files, the sub-files are grouped  into $K$ disjoint and equal-size subsets and each user caches the sub-files in a different subset which corresponds to caching at level 1.\\
\indent Now, to analyze the delivery phase with cross-level coded delivery, let $\mathcal{P}(\mathcal{K})$ be the power set of $\mathcal{K}$, and $\mathbb{S}\in\mathcal{P}(\mathcal{K})$ be the set of sets $\mathcal{S}$ with cardinality $t+1$, i.e., $\vert \mathbb{S} \vert={K \choose t+1}$, and $\tilde{\mathbb{S}} \subseteq \mathbb{S}$ defined as $\tilde{\mathbb{S}}=\left\{ \mathcal{S}\in \mathbb{S}:\mathcal{K}^{h}\cap \mathcal{S} \neq 0 \right\}$. Recall that, with the conventional coded delivery scheme, the low-level and high-level users are served separately. Hence, for the delivery of the  high-level files, for each set $\mathcal{S} \subseteq \tilde{\mathbb{S}}$ the server transmits
\begin{equation}\label{convent}
\oplus_{i\in \mathcal{K}^{h}\cap \mathcal{S}}W_{d_{i},\mathcal{S}\setminus\left\{i\right\}}
\end{equation}
in order to deliver all the sub-files requested by the high-level users. However, although the high-level files are cached at level $t$, a message destined for high-level users has a multicasting gain of $\mu_{\mathcal{S}}=\vert \mathcal{K}^{h}\cap \mathcal{S}\vert$, instead of $t+1$. In particular, when 
$\mu_{\mathcal{S}}=1$ the corresponding message is a unicast message.\\
\indent We call a multicast message in the form of (\ref{convent}) as {\em decomposable} if in the corresponding set $\mathcal{S}$, there is at least one high-level and one low-level user. The server can utilize those high-level sub-files for cross-level delivery. Let $\mathbb{S}_{d}$ be the set of sets $\mathcal{S}$ that correspond to a decomposable multicast message. In the $\mathrm{CL(t_h,1,N_h, N_l, N_r)}$ scheme, for each demand realization  $\mathbf{k}$, we look for a set $\mathbb{S}_{cl}\subseteq \mathbb{S}_{d}$ to use for pairing sub-files of high-level users and sub-files of low-level users.
We introduce variable $x_{\mathcal{S}}\in\left\{1,0\right\}$, where $x_{\mathcal{S}}=0$ if the conventional coded delivery scheme, as given in (\ref{convent}), is used to deliver the sub-files corresponding to set $\mathcal{S}$, while $x_{\mathcal{S}}=1$ if the corresponding multicast message is decomposed and the sub-files in the message are delivered by cross-level coded delivery. We further introduce variables $x_{(\mathcal{S},i,j)}$, for $i\in K^{h}\cap \mathcal{S}$ and $j\in K^{l}\cap \mathcal{S}$, where $x_{(\mathcal{S},i,j)}=1$, if sub-file $W_{d_{i},\mathcal{S}\setminus\left\{i\right\}}$ is paired with a sub-file $W_{d_{j},\left\{i,***\right\}}$ that is requested by low-level user $j$ and cached by high-level user $i$. Then, the server multicasts $W_{d_{i},S\setminus\left\{i\right\}} \oplus W_{d_{j},\left\{i,***\right\}}$. Hence, for each set $\mathcal{S}$ there are $\mu_{\mathcal{S}} \times \vert \mathcal{K}^{l}\cap \mathcal{S}\vert$  variables in the form $x_{(\mathcal{S},i,j)}$.\\
\indent To illustrate the multicast message decomposition process, consider the $\mathrm{CL(t_h=3,1,N_h, N_l, N_r)}$ scheme, and assume that there are $K=7$ users  with a demand realization $\mathbf{k}$,
 $\mathcal{K}^{h}=\left\{1,2,3,4\right\}$ and $\mathcal{K}^{l}=\left\{5,6,7\right\}$. Consider a particular set $\mathcal{S}=\left\{1, 2, 6, 7\right\}$. According to (\ref{convent}), the server multicasts $W_{d_{1},\left\{2,6,7\right\}} \oplus W_{d_{2},\left\{1,6,7\right\}}$ for high-level users 1 and 2. With a multicast message decomposition specified by $x_{(\mathcal{S},1,6)}=1$, $x_{(\mathcal{S},1,7)}=0$, $x_{(\mathcal{S},2,6)}=0$, $x_{(\mathcal{S},2,7)}=1$, and $x_{\mathcal{S}}=1$, the server multicasts the  messages  $W_{d_{6},\left\{1,***\right\}}\oplus W_{d_{1},\left\{2,6,7\right\}}$ and $W_{d_{7},\left\{2,***\right\}}\oplus W_{d_{2},\left\{1,6,7\right\}}$ in the cross-level delivery phase. We remark that, if the conventional coded delivery scheme is used, sub-files $W_{d_{6},\left\{1,***\right\}}$ and $W_{d_{7},\left\{2,***\right\}}$ are sent as  unicast messages. Hence, in total three messages would be transmitted with the conventional coded delivery scheme, i.e., $W_{d_{1},\left\{2,6,7\right\}} \oplus W_{d_{2},\left\{1,6,7\right\}}$, $W_{d_{6},\left\{1,***\right\}}$, and $W_{d_{7},\left\{2,***\right\}}$; however, via CLCD this can be reduced to two, i.e., $W_{d_{6},\left\{1,***\right\}}\oplus W_{d_{1},\left\{2,6,7\right\}}$ and $W_{d_{7},\left\{2,***\right\}}\oplus W_{d_{2},\left\{1,6,7\right\}}$. One can observe that if a multicast message is decomposed and used by the CLCD scheme, then the number of multicast messages is reduced by one.\\
\indent Note that variables $x_{(\mathcal{S},i,j)}$ and $x_{(\mathcal{S},i,k)}$ cannot be 1 at the same time, since the sub-file $W_{d_{i},\mathcal{S}\setminus\left\{i\right\}}$ can be paired with only one low-level sub-file. Hence, for the message decomposition we have the following constraint
\begin{equation}
\sum_{j\in \mathcal{K}^{l}\cap \mathcal{S}} x_{(\mathcal{S},i,j)} \leq 1,~~ \forall i\in \mathcal{K}^{h}\cap \mathcal{S}\text{ and } \forall\mathcal{S}\in\mathbb{S}_{d}.
\end{equation}
 Another constraint, from the construction, is that each high-level user $i$ contains ${K-1 \choose t-1}/t$ sub-files that are missing at low-level user $j$; thus, the total number of sub-file pairings between a low-level user $j$ and a high-level user $i$ is at most ${K-1 \choose t-1}/t$ i.e.,
\begin{equation}
\sum_{\mathcal{}S\in\mathbb{S}_{d}} x_{(\mathcal{S},i,j)} \leq \frac{ {K-1 \choose t-1}}{t}~~ \forall i\in\mathcal{K}^{h},j\in\mathcal{K}^{l}.
\end{equation}
Finally, if $x_{S}=1$, then all the corresponding sub-files must be sent via CLCD,  i.e.,
\begin{equation}
\sum_{i\in \mathcal{K}^{h}\cap \mathcal{S}, j\in \mathcal{K}^{l}\cap \mathcal{S}} x_{(\mathcal{S},i,j)} = \mu_{\mathcal{S}} x_{S}, ~~ \forall \mathcal{S}\in\mathbb{S}_{d}.
\end{equation}
\begin{algorithm}[t]\footnotesize{
		\For{$\mathcal{S}\in\tilde{\mathbb{S}}\setminus\mathbb{S}_{cl}$}{
                     multicast $\oplus_{i\in \mathcal{K}^{h}\cap \mathcal{S}}W_{d_{i},\mathcal{S}\setminus\left\{i\right\}}$\;
											   }										
	  \For{$\mathcal{S}\in\mathbb{S}_{cl}$}{	
		                    \For{$i\in \mathcal{K}^{h}\cap \mathcal{S}$}{
                                                 \For{$j\in \mathcal{K}^{l}\cap \mathcal{S}$}{
												             \If{$x_{(\mathcal{S},i,j)}=1$}{
								              multicast $W_{d_{i},\mathcal{S}\setminus\left\{i\right\}} \oplus W_{d_{j},\left\{i,***\right\}}$\;
											                                               } 
		                                                                                     }
											                           }
											         }
    	\For{$j,k\in\mathcal{K}^{l}$}{
                    send $W_{d_{k},\left\{j,***\right\}} \oplus W_{d_{j},\left\{k,***\right\}}$\;
											   }
			Remaining sub-packets are send via unicast transmission\;											}									\caption{Overall Delivery phase of $\mathrm{CL(t_{1},1,N_{h},N_{l},N_{r})}$}
\label{alg:clt10}
\end{algorithm}
For the given values of the variables $x_{(\mathcal{S},i,j)}$ and $x_{\mathcal{S}}$, the delivery phase of $\mathrm{CL(t_{1},1,N_{h},N_{l},N_{r})}$  is given in Algorithm \ref{alg:clt10}. According to Algorithm \ref{alg:clt10}, the number of transmitted messages, $N_{T}(\mathbf{k})$, for a given demand realization $\mathbf{k}=\left[k^{h},k^{l},k^{r}\right]$ is given by
\begin{align}
N_{T}(\mathbf{k})=&{K\choose t+1}-{k^{r}+k^{l}\choose t+1}-\sum_{\mathcal{S}\in\mathbb{S}_{d}}x_{\mathcal{S}}\\
&+\frac{{K-1 \choose t-1}}{t}k^{h}k^{l}+\frac{{K-1 \choose t-1}}{t}k^{r}k^{l}+{k^{l}\choose 2}\frac{{K-1 \choose t-1}}{t}.
\end{align}
Hence, the main objective of $\mathrm{CL(t_h,t_l=1,N_h, N_l, N_r)}$ scheme is to maximize $\sum_{\mathcal{S}\in\mathbb{S}_{d}}x_{\mathcal{S}}$ in order to minimize the delivery rate for each possible demand realization. Equivalently, the minimum delivery rate problem can be converted to the following optimization problem: 
\begin{small}
\begin{align}
   \text{\bf P2:} \;\;\;\max&
   \begin{aligned}[t]
       \sum_{\mathcal{S}\in\mathbb{S}_{d}} x_{\mathcal{S}}  \notag
   \end{aligned} \\
  \text{subject to: }  \sum_{j\in \mathcal{K}^{l}\cap \mathcal{S}} x_{(\mathcal{S},i,j)} &\leq 1, ~~~ \forall \mathcal{S}\in\mathbb{S}_{d},\\
  \sum_{\mathcal{S}\in\mathbb{S}_{d}} x_{(\mathcal{S},i,j)}  &\leq {K-1 \choose t-1}/t, ~~~ \forall i\in \mathcal{K}^{h},j\in \mathcal{K}^{l},\\
	 \sum_{i\in \mathcal{K}^{h}\cap \mathcal{S}, \forall j\in \mathcal{K}^{l}\cap \mathcal{S}} x_{(\mathcal{S},i,j)}& = \mu_{\mathcal{S}}x_{\mathcal{S}}, ~~~ \forall \mathcal{S}\in\mathbb{S}_{d}.
\end{align}
\end{small}
By solving   \text{\bf P2} for each demand realization $\mathbf{k}$, we can construct the optimal coded delivery scheme for $\mathrm{CL(t_h,1,N_h, N_l, N_r)}$. We remark that \text{\bf P2} is a {\em binary integer programming} problem, and unlike $\mathrm{CL(2,1,N_h, N_l, N_r)}$, the optimization of the delivery phase cannot be solved in polynomial time. Nevertheless,  complexity of  \text{\bf P2} does not depend on the number of files, $N$, but depends on the number of users, $K$; hence, for moderate number of users the CLCD  scheme can be optimized even for a very large file library. Once the required delivery rate for each possible demand realization is computed, optimal placement strategy; that is, the number of files not to cache, can be found following the same procedure in Section \ref{s:CLCD}. We want to remark that instead of solving \text{\bf P2}, it is also possible to  design a greedy algorithm to construct the multicast messages.\\
\indent In Section \ref{s:clcd_ext2}, we will study the most general  scheme $\mathrm{CL(t_h,t_l,N_h, N_l, N_r)}$ and show that the delivery phase can be still analyzed as an optimization problem with a reduced computational complexity. But before extending our analysis, we would like to present the numerical results comparing the CLCD schemes presented so far with the alternative approaches in the literature.

\begin{figure*}
    \centering
         \begin{subfigure}[b]{0.45\textwidth}
        \includegraphics[scale=0.6]{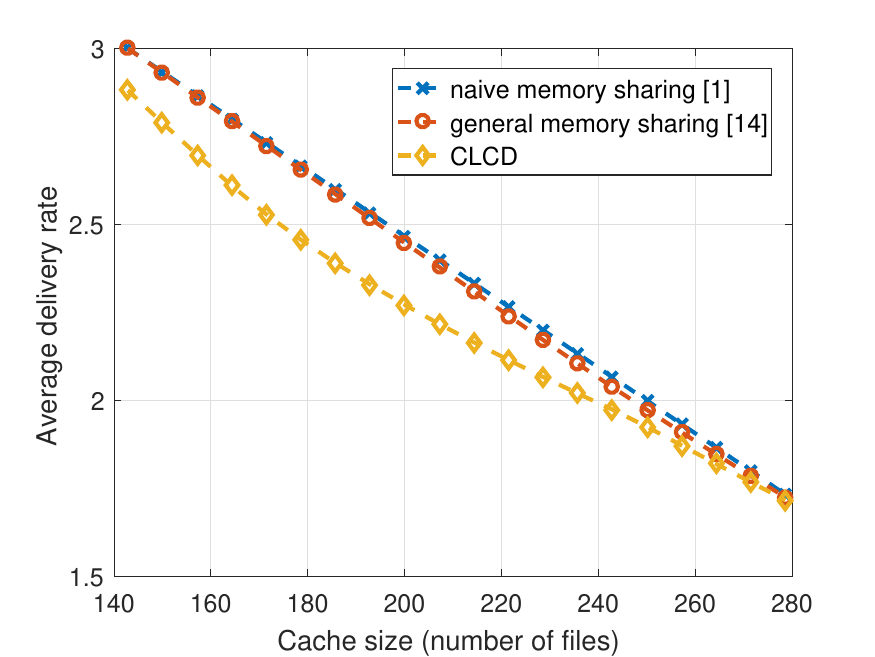}
        \caption{$\gamma=0.7$.}
    \end{subfigure}
    \begin{subfigure}[b]{0.45\textwidth}
        \includegraphics[scale=0.6]{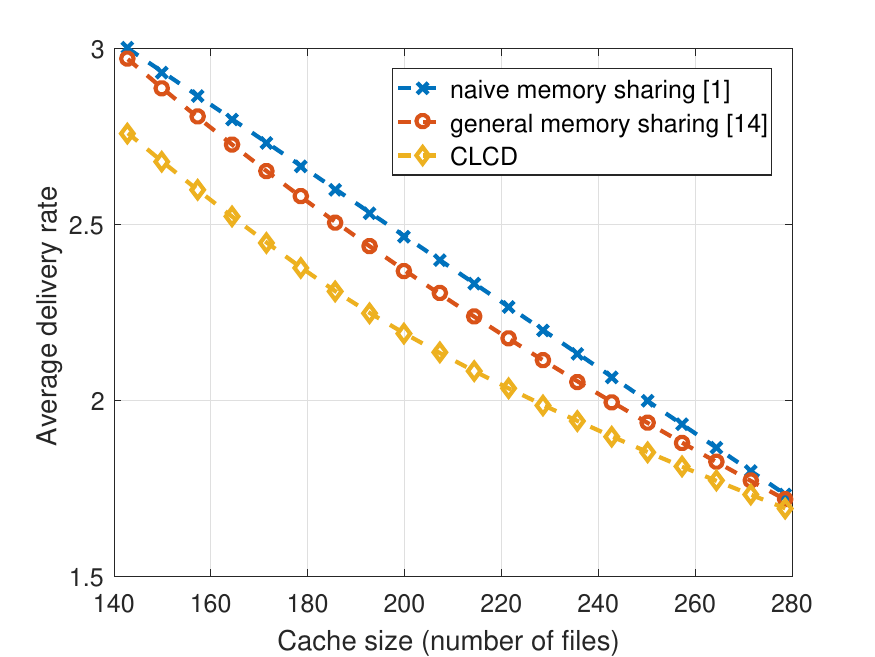}
        \caption{$\gamma=0.75$.}
        \end{subfigure}
    \begin{subfigure}[b]{0.45\textwidth}
        \includegraphics[scale=0.6]{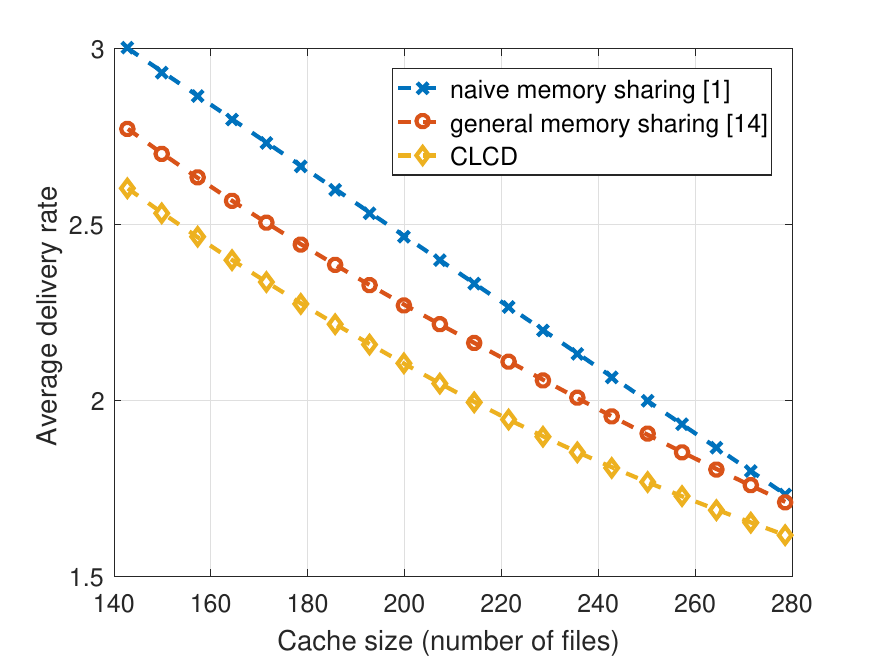}
        \caption{$\gamma=0.8$.}
    \end{subfigure}
      \begin{subfigure}[b]{0.45\textwidth}
        \includegraphics[scale=0.6]{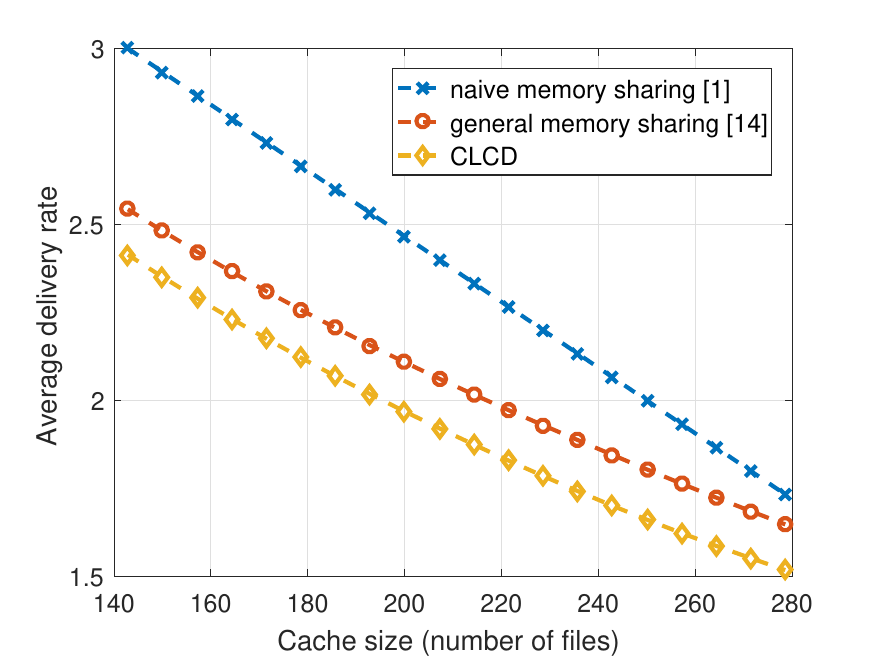}
        \caption{$\gamma=0.85$.}
    \end{subfigure}
    \caption{Average delivery rate vs. the normalized cache size $M$ for different Zipf parameter ($\gamma$) values.}
		\label{results1}
\end{figure*}
\begin{figure*}
    \centering
         \begin{subfigure}[b]{0.45\textwidth}
        \includegraphics[scale=0.6]{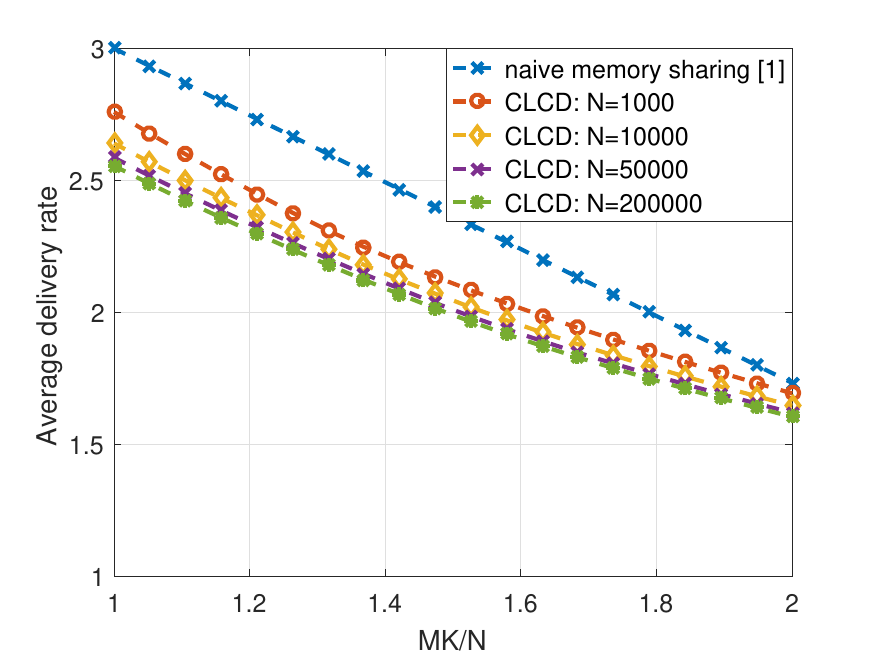}
        \caption{$\gamma=0.75$.}
    \end{subfigure}
    \begin{subfigure}[b]{0.45\textwidth}
        \includegraphics[scale=0.6]{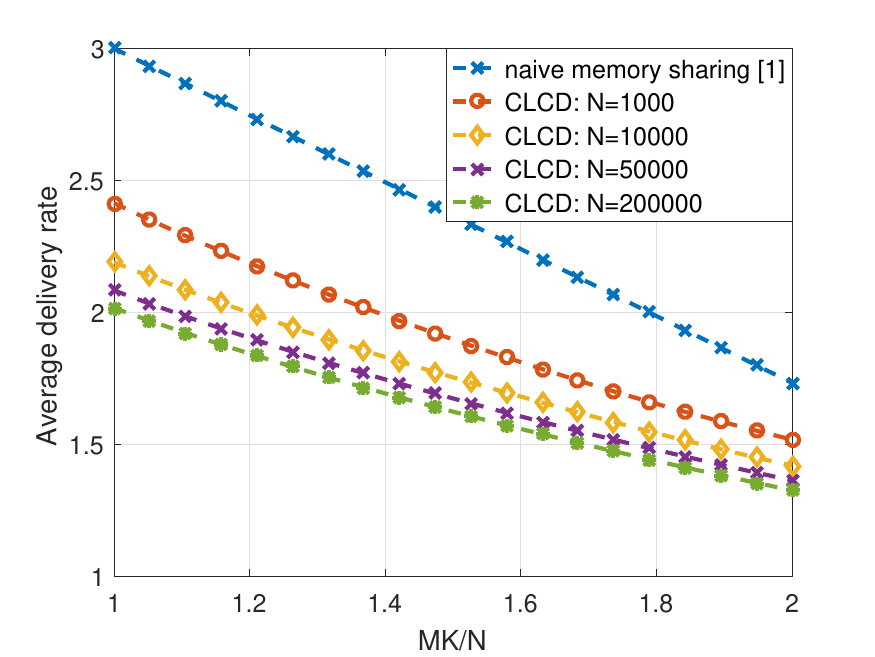}
        \caption{$\gamma=0.85$.}
        \end{subfigure}
    \caption{Average delivery rate vs. $MK/N$ for different Zipf parameter ($\gamma$) values.}
		\label{results2}
\end{figure*}
\section{Numerical Results}\label{s:sim}
In this section, we compare the performance of the proposed CLCD scheme with that of the conventional centralized coded delivery scheme with two different content placement strategies. The first content placement strategy is called {\em naive memory sharing}, introduced in \cite{CD.F1}, in which all the files are cached at the same level according to a single parameter $t=MK/N$. Note that, when parameter $t$ is not an integer, then the  files in the library are divided into two disjoint fragments identically, and these fragments are cached according to parameters $\left\lfloor t\right\rfloor$ and $ \left\lceil t\right\rceil$.\\
\indent The second benchmark strategy is the {\em general memory sharing} scheme proposed in \cite{CD.ND6}, which is shown to outperform other coded delivery techniques under non-uniform demand distributions. In this scheme, each file is divided into $K+1$ disjoint fragments, and the $k$th fragment, $1<k$, is cached according to parameter $t=k-1$, while the first fragment is not cached. Thus, the overall system can be considered as a combination of $K+1$ sequential coded delivery phases with different multicasting gains, i.e.,  the $k$th delivery phase is executed with multicasting gain of  $k+1$. The size of  each fragment of  each file can be obtained via solving a linear optimization problem \cite{CD.ND6}.\\ 
\indent In general, it has been observed that the popularity of video files for on-demand streaming applications approximately follows a Zipf distribution with parameter $ 1>\gamma> 0.65$ \cite{VS1}, \cite{VS2}. In our simulations, we consider $\gamma=0.7, 0.75, 0.8, 0.85$. We note that  $\gamma$, represents the skewness of the distribution of video popularity, where smaller $\gamma$ values indicates a more uniform popularity distribution. In realistic scenarios, number of files in the video library is considered to be on the order of $10^4$. However, due to the complexity of the general memory sharing scheme, we will consider $N=1000$ and $K=7$ in our initial simulations. We let the  cache size $M$ vary from 140 to 280, which corresponds to $1<t<2$. We are particularly interested in this regime as all three strategies considered in this paper converge to the same performance for larger $t$ values. The delivery rates achieved by the naive memory sharing, the general memory sharing, and the CLCD schemes are illustrated in Figure \ref{results1}. We note that, for each value of the Zipf parameter $\gamma$ we calculate the achievable delivery rate of the scheme $\mathrm{CL(t_{h},1,N_{h},N_{l},N_{r})}$, for $t_{h}=2,3,\ldots,6$, and  take the minimum of them.\\
\indent We observe that when the number of files $N$ is  large, the general memory sharing algorithm tends to divide the library  into two groups according popularities, such that the most popular files  are cached according to the naive memory sharing scheme while the less popular files are not cached at all. One can observe from Fig. \ref{results1} that when $\gamma=0.7$ or $\gamma=0.75$ the optimal memory sharing scheme performs very similarly to the naive memory sharing scheme, while the proposed CLCD scheme can provide a significant reduction in the achievable delivery rate. In addition, we observe that CLCD outperforms both the naiive and general memory sharing schemes at all $\gamma$ values considered here.\\
\indent We also perform simulations to illustrate how the number of files $N$ affects the performance of the CLCD scheme when $t=MK/N$ is fixed, i.e., the cache memory size scales with the number of files in the library. In the naive memory sharing scheme, all the files are cached at level $t=MK/N$, thus the performance of this scheme does not change with the size of the file library as long as $M/N$ is fixed. In the simulations, we consider $N=10^{3},10^{4},5\times10^{4}$, and $2\times10^{5}$. Due to its complexity it is not possible to run  the general memory sharing scheme for large libraries, thus we compare CLCD scheme only with the naive memory sharing scheme. We observe in Fig. \ref{results2} that the CLCD scheme performs better for  larger file libraries. In particular, when $\gamma=0.85$ the performance gap between the CLCD scheme and the naive memory sharing scheme for $N=2\times10^{5}$ is almost two times of the performance gap for $N=10^{3}$. 
\section{ General $\mathrm{CL(t_h,t_l,N_h, N_l, N_r)}$ Scheme}\label{s:clcd_ext2}
Having shown the superiority of the proposed $\mathrm{CL(t_h,1,N_h, N_l, N_r)}$ scheme to its alternatives in the literature, in this section, we present the most generic form of the CLCD scheme. In the previously introduced versions with $t_l=1$, the placement phase is designed to prevent any miss-alignment between the  high-level and low-level sub-file sizes. Hence, the fundamental modification in the $\mathrm{CL(t_h,t_l,N_h, N_l, N_r)}$ scheme with respect to the previous versions is that the alignment of the sub-file sizes and the construction of the multicast messages are done jointly during the deliver phase. In this section, for sake of clarity, we start our presentation by setting $N_r=0$. This constraint will later be removed. The general structure of the $\mathrm{CL(t_h,t_l,N_h, N_l, 0)}$ scheme is given in Algorithm \ref{alg:CLCD_general}.

\begin{algorithm}[t]{\footnotesize
    \SetKwInOut{Input}{Input}
    \SetKwInOut{Output}{Output}
    \Input{$\mathcal{K}^{h}$, $\mathcal{K}^{l}$}
		\For{all $\mathcal{S} \subseteq \mathcal{K}$ :  $\vert \mathcal{S}\vert =t_{h}+1$}{ 
         \If{ $\mathcal{S} \subseteq \mathcal{K}^{h}$}{ serve high-level users with multicasting gain $t_h +1$} 
         \If{ $\mathcal{S} \nsubseteq \mathcal{K}^{h}$ and $\mathcal{S} \nsubseteq \mathcal{K}^{l}$}{ \For{all $\hat{\mathcal{S}}\subseteq\mathcal{S}$ :  $\vert \hat{\mathcal{S}}\vert =t_l +1$ and $\hat{\mathcal{S}}\nsubseteq \mathcal{K}^{l}$}{ use CLCD with multicasting gain of $t_l +1$} 
                                                        } 
											  }
		\For{all $\mathcal{S} \subseteq \mathcal{K}^{l}$ :  $\vert \mathcal{S}\vert =t_l+1$}{ serve low-level users with multicasting gain of $t_l +1$}
}
\caption{$\mathrm{CL(t_h,t_l,N_h, N_l, N_r=0)}$ scheme}
\label{alg:CLCD_general}
\end{algorithm}
Here, the main design issue is how to overlap the bits of high-level and low-level files that are delivered in the CLCD phase. The key point here is that, for a subset $\mathcal{S}$ of different types of users, i.e., $\mathcal{S} \nsubseteq \mathcal{K}^{h}$ and $\mathcal{S} \nsubseteq \mathcal{K}^{l}$ with  $\vert \mathcal{S}\vert =t_{h}+1$, any high level user $i\in\mathcal{S}$ appears in exactly $t_h\choose t_l$ of the subsets $\hat{\mathcal{S}}$ of $\mathcal{S}$ with $\vert\hat{\mathcal{S}}\vert=t_{l}+1$. Hence, the requested sub-file $W_{d_{i},\mathcal{S}\setminus\{i\}}$ should be divided into $t_h\choose t_l$ smaller sub-files. Similarly, from the perspective of low-level users, each $\hat{\mathcal{S}}$ appears as a subset of exactly $K-(t_l+1)\choose t_h-t_l$ different $\mathcal{S}$ sets, thus the sub-file requested  by low-level user $j$, $W_{d_{j},\hat{\mathcal{S}}\setminus\{j\}}$, should be divided into $K-(t_l+1)\choose t_h-t_l$ smaller sub-files. Let $F^h_{cl}$ and $F^l_{cl}$ be the sizes of the sub-files used for the aforementioned high-level and low-level sub-files, respectively, in the cross-delivery phase, i.e.,
\begin{equation}
F^h_{cl} = \left({K \choose t_h} {t_h \choose t_l}\right)^{-1},
\label{hfs}
\end{equation}
and
\begin{equation}
F^l_{cl} = \left({K \choose t_l} {K-(t_l+1) \choose t_h-t_l}\right)^{-1}.
\label{hfl}
\end{equation}
By expanding (\ref{hfs}) and (\ref{hfl}) one can observe that 
\begin{equation}
F^l_{cl} = F^h_{cl} \times \frac{K-t_l}{K-t_h},
\label{file_gap}
\end{equation}
which implies that it is possible to overlap the high-level files with the low-level files in the cross-delivery step as long as there is at least one low-level user in $\hat{\mathcal{S}}$. Accordingly, the overall rate, for given $k^{h},k^{l}$ can be written as 
\begin{align}
R(k^{h},k^{l})=&{k^{h}\choose t_h+1}F_h + \left({K\choose t_l+1}-{k^{h}\choose t_l+1}\right)F_l \label{clcd_rate}\\ 
&+ \underbrace{{k^{h}\choose t_l+1}\left({K-(t_l +1)\choose t_h-t_l}-{k^{h}-(t_l +1)\choose t_h-t_l}\right)F^h_{cl}.}_{R_{no}:\text{ non-overlapped bits of high level files}}\nonumber
\end{align}
\indent Here, the CLCD scheme has two objectives: one is to achieve a multicasting gain of at least $t_l+1$ in the delivery phase, and the other is to overlap the delivery of the high-level and low-level files as much as possible. Hence, we want to minimize the term  $R_{no}$ in (\ref{clcd_rate}). This can be achieved by dividing the high-level files in a non-uniform manner. To clarify the non-uniform file division, consider the case with $t_h=4$, $t_l=3$, and assume that there are $K=7$ users  with a demand realization $\mathbf{k}=[4,3,0]$, where  $\mathcal{K}^{h}=\left\{1,2,3,4\right\}$ and $\mathcal{K}^{l}=\left\{5,6,7\right\}$. \\
 \indent According to the proposed CLCD design, for set $\mathcal{S}=\left\{1,2,3,4,5\right\}$, delivery of the requested sub-files are realized over 5 steps, with a multicasting gain of $t_l+1=4$, corresponding to the subsets $\hat{\mathcal{S}}_{1}=\left\{1,2,3,4\right\}$, $\hat{\mathcal{S}}_{2}=\left\{1,2,3,5\right\}$, $\hat{\mathcal{S}}_{3}=\left\{1,2,4,5\right\}$, $\hat{\mathcal{S}}_{4}=\left\{1,3,4,5\right\}$, $\hat{\mathcal{S}}_{5}=\left\{2,3,4,5\right\}$. We recall that since user 1 receives its required file $W_{d_1,\left\{2,3,4,5\right\}}$, part by part in four of these steps, the corresponding sub-file is further divided into $4$ smaller sub-files. Let $W_{d_1,\left\{2,3,4,5\right\},\hat{\mathcal{S}}_{j}}$ denote the one  delivered in the step corresponding to subset $\hat{\mathcal{S}}_{j}$. Subset $\hat{\mathcal{S}}_{1}=\left\{1,2,3,4\right\}$ consists of only high-level users. We also recall that, as given in (\ref{file_gap}), the low-level sub-files are larger than the high-level ones. Hence, to reduce the amount of non overlapping bits $R_{no}$, one can reduce the size of $W_{d_1,\left\{2,3,4,5\right\},\hat{\mathcal{S}}_{1}}$ and, accordingly, increase the size of $W_{d_1,\left\{2,3,4,5\right\},\hat{\mathcal{S}}_{2}}$, $W_{d_1,\left\{2,3,4,5\right\},\hat{\mathcal{S}}_{3}}$,$W_{d_1,\left\{2,3,4,5\right\},\hat{\mathcal{S}}_{4}}$ until they match with that of $F^l_{cl}$.\\
 \indent The main question arising from this observation is how to adjust the size of the high-level sub-files assigned to each $\hat{\mathcal{S}}$ so that the high-level and low-level sub-files overlap as much as possible. Let us focus on a particular subset $\mathcal{S}\nsubseteq\mathcal{K}^{h}$, $\mathcal{S}\nsubseteq\mathcal{K}^{l}$ and a particular high-level user $i\in\mathcal{S}$. Let the number of subsets $\hat{\mathcal{S}}:i\in\hat{\mathcal{S}}\subseteq\mathcal{K}^{h}$, be denoted by $n^{i}_h(\mathcal{S})$, and the number of subsets $\hat{\mathcal{S}}:i\in\hat{\mathcal{S}}\nsubseteq\mathcal{K}^{h}$ be denoted by $n^{i}_{cl}(\mathcal{S})$. Since we want to reduce the size of sub-files $W_{d_i,\mathcal{S}\setminus\left\{i\right\},\hat{\mathcal{S}}}$, $i\in\hat{\mathcal{S}}\subseteq\mathcal{K}^{h}$, and increase that of sub-files $W_{d_i,\mathcal{S}\setminus\left\{i\right\},\hat{\mathcal{S}}}$,  $i\in\hat{\mathcal{S}}\nsubseteq\mathcal{K}^{h}$, we have the following constraint based on (\ref{file_gap}):
 \begin{equation}
 \left(\frac{n^{i}_h(\mathcal{S})\alpha_{i}(\mathcal{S})}{n^{i}_{cl}(\mathcal{S})}+1\right)F^h_{cl}=F^h_{cl}\frac{K-t_l}{K-t_h},
 \end{equation}
where $\alpha_{i}(\mathcal{S})$ denotes the ratio of reduction in the size of the high-level sub-files. Accordingly, $\alpha_{i}(\mathcal{S})$ parameter can be written as 
\begin{equation}
 \alpha_{i}(\mathcal{S})=\frac{(t_h-t_l)n^{i}_{cl}(\mathcal{S})}{(K-t_h)n^{i}_{h}(\mathcal{S})}.
\end{equation}
We note that, parameters $n^{i}_{cl}(\mathcal{S})$, $n^{i}_h(\mathcal{S})$ and $\alpha_i(\mathcal{S})$ are identical for all the high-level users $i\in\mathcal{S}$; hence, we drop the user index for simplicity. Once the size of the high-level sub-files are adjusted, $R_{no}$ can be rewritten as follows:
\begin{equation}
R_{no}=\sum_{\mathcal{S}\subseteq \mathcal{K}:~ \mathcal{S}\cap \mathcal{K}^h\neq\emptyset,~\mathcal{S}\cap \mathcal{K}^l\neq\emptyset}n_h(\mathcal{S})\min(1-\alpha(\mathcal{S}),0)F^h_{cl}.
\label{min_overlap}
\end{equation}
Consequently, by inserting $R_{no}$  above into  (\ref{clcd_rate}), one can obtain the required delivery rate for Algorithm \ref{alg:CLCD_general}.\\
\indent We remark that with the CLCD scheme all the sub-files are served with a minimum multicasting gain of $t_l+1$. If $R_{no}=0$, one can claim the optimality of the delivery phase, under the same placement scheme, since the low-level files cannot be served with a multicasting gain greater than $t_l+1$. On the other hand, if $R_{no} > 0$, then the non-overlapped bits of the high-level sub-files are delivered with a multicasting gain of $t_l + 1$, although, in principle, they could be delivered with a multicasting gain of $t + 1$, where $t_h > t > t_l$; thus, when $ t_h >> t_l $ this approach might be inefficient. Hence, the main question at this point is how to decide which $\mathcal{S}$ subsets with $\vert\mathcal{S}\vert=t_h+1$ will be used for CLCD. To this end, we introduce another parameter, which denoted by $t_{th}$, where $t_h > t_{th} \geq t_l$, such that $\mathcal{S}:\vert\mathcal{S}\vert=t_h+1$ is used for CLCD if $\vert\mathcal{S}\cap\mathcal{K}^{h}\vert \leq t_{th}+1$, otherwise high-level users in $\mathcal{S}$ are served with multicasting gain of $\vert\mathcal{S}\cap\mathcal{K}^{h}\vert$ as in Algorithm \ref{alg:CLCD_general}.

The CLCD scheme with $t_{th}$ is given in Algorithm \ref{alg:clcd_th}.
\begin{algorithm}[t]{\footnotesize
    \SetKwInOut{Input}{Input}
    \SetKwInOut{Output}{Output}
    \Input{$\mathcal{K}^{h}$, $\mathcal{K}^{l}$, $t_{th}$}
    Construct set $\mathbb{S}_{cl}=\left\{\mathcal{S}:\vert\mathcal{S}\cap\mathcal{K}^{h}\vert \leq t_{th}+1, ~~ \mathcal{S}\nsubseteq\mathcal{K}^{l}\right\}$\;
    Based on $\mathbb{S}_{cl}$ compute initial values of $F^{l}_{cl}$ and $F^{h}_{cl}$\;
    \For{all $\mathcal{S} \subseteq \mathbb{S}_{cl}$}{Compute $n_h(\mathcal{S})$, $n_l(\mathcal{S})$, $\alpha(\mathcal{S})$ to seek maximum overlapping}
		\For{all $\mathcal{S} \subseteq \mathcal{K}$ :  $\vert \mathcal{S}\vert =t_{h}+1$}{ 
         \If{ $\vert \mathcal{S}\cap\mathcal{K}^{h}\vert>  t_{th}+1$}{ serve high-level users with multicasting gain  $\vert \mathcal{S}\cap\mathcal{K}^{h}\vert$} 
         \If{ $\mathcal{S} \subseteq \mathbb{S}_{cl}$}{ \For{all $\hat{\mathcal{S}}\subseteq\mathcal{S}$ :  $\vert \hat{\mathcal{S}}\vert =t_l +1$ and $\hat{\mathcal{S}}\nsubseteq \mathcal{K}^{l}$}{ use CLCD with multicasting gain $t_l +1$} 
                                                        } 
											  }
		\For{all $\mathcal{S} \subseteq \mathcal{K}^{l}$ :  $\vert \mathcal{S}\vert =t_l+1$}{ serve low-level users with multicasting gain of $t_l +1$}
}
\caption{$\mathrm{CL(t_h,t_l,N_h, N_l, N_r=0)}$ scheme with threshold $t_{th}$}
\label{alg:clcd_th}
\end{algorithm}
The delivery rate, $R(\mathbf{k})$, for given demand realization $\mathbf{k}$ with Algorithm \ref{alg:clcd_th} can be written as,
\begin{align}
R(\mathbf{k})=&\sum_{\mathcal{S}\subseteq \mathcal{K}:~ \vert\mathcal{S}\cap \mathcal{K}^h\vert>t_{th}+1}F_h + \left({K\choose t_l+1}-{k^h\choose t_l+1}\right)F_l\label{final_rate}\\ 
&+ \sum_{\mathcal{S}\subseteq\mathbb{S}_{cl}}n_h(\mathcal{S})\min(1-\alpha(\mathcal{S}),0)F^h_{cl}.\nonumber
\end{align}
The second term on the right hand side of  (\ref{final_rate}) does not depend on  parameter $t_{th}$; whereas the first term decreases with $t_{th}$, while the last term increases, which implies that $t_{th}$ seeks a balance between the first and the last terms. Therefore, for each  $\mathbf{k}$, a different $t_{th}$ value may minimize $R(\mathbf{k})$. The average delivery rate is given by
\begin{equation}
\bar{R}=\sum_{\mathbf{k}}P(\mathbf{k})R(\mathbf{k}).
\label{exp_rate_th}
\end{equation}
\indent In general, given $t_{h}$ and $t_{l}$, one should find the best $t_{th}$ value for each $\mathbf{k}$.\\
\indent In the most generic form of the CLCD scheme we also allow $N_r>0$. We slightly modify the delivery scheme by simply considering the zero-level users (users with uncached file requests) as low-level users for cross-level delivery and serve them with unicast transmissions at the end. Accordingly, the delivery rate is given by as
\begin{align}
R(\mathbf{k})=&\sum_{\mathcal{S}\subseteq \mathcal{K}:~ \vert\mathcal{S}\cap \mathcal{K}^h\vert>t_{th}+1}F_h\nonumber\\  &+ \left({K\choose t_l+1}-{k^h\choose t_l+1}-{k^r\choose t_l+1}\right)F_l\label{final_rate2}\\ 
&+ \sum_{\mathcal{S}\subseteq\mathbb{S}_{cl}}n_h(\mathcal{S})\min(1-\alpha(\mathcal{S}),0)F^h_{cl} +k^{r}.\nonumber
\end{align}
We note that it might be possible to further reduce the delivery rate  by revisiting the sub-file size alignment strategy  taking into account the zero-level users at the expense of additional complexity.

\section{Conclusions}\label{s:conclusion}
 We introduced a novel centralized coded caching delivery scheme, called  CLCD,  for cache-aided content delivery under non-uniform demand distributions. The proposed caching scheme uses a different placement strategy for the files depending on their popularities, such that the cache memory allocated to the sub-files belonging to more popular files are larger compared to the low popular ones.
The main novelty of the proposed approach is in the delivery phase, where the messages multicasted to the group of users are carefully designed to satisfy as many users as possible with minimal delivery rate while in the case where the  files are cached in a non-uniform manner based on their popularities.
For a certain special case of the proposed algorithm we are able to provide a closed form expression for the delivery rate. We also showed via numerical simulations that the proposed CLCD scheme can provide up to $10 \%$ reduction in the average delivery rate compared to the state-of-the-art. We expect this gain to grow considerably as the number of files increases, while the state-of-the-art schemes in the literature cannot be implemented for a large file library due to their formidable complexity.

\appendices\label{appendix}
\section{Set partitioning}\label{appendix_partition}
In this section, we will introduce some useful lemmas and definitions related to graph theory that will be utilized in the process of multicast message construction, particularly for pairing the sub-files. We use $G$, $\mathcal{V}(G)$ and $\mathcal{E}(G)$ to denote a graph, its set of vertices and edges, respectively. For simplicity, we enumerate the vertices, and set $\mathcal{V}(G)=[V]$, so that $V$ represents the number of vertices.

\begin{definition}
A graph $G$ is a complete graph if each pair of distinct vertices is connected by a unique edge. 
\end{definition}

\begin{definition}
A matching $D$ of graph $G$ is a subgraph of $G$ whose edges share no vertex; that is, each vertex in matching $D$ has degree one.   
\end{definition}

\begin{definition}
A matching is maximum when it has the largest possible size.  
\end{definition}

\begin{figure*}
    \centering
    \begin{subfigure}[b]{0.3\textwidth}
        \includegraphics[scale=0.2]{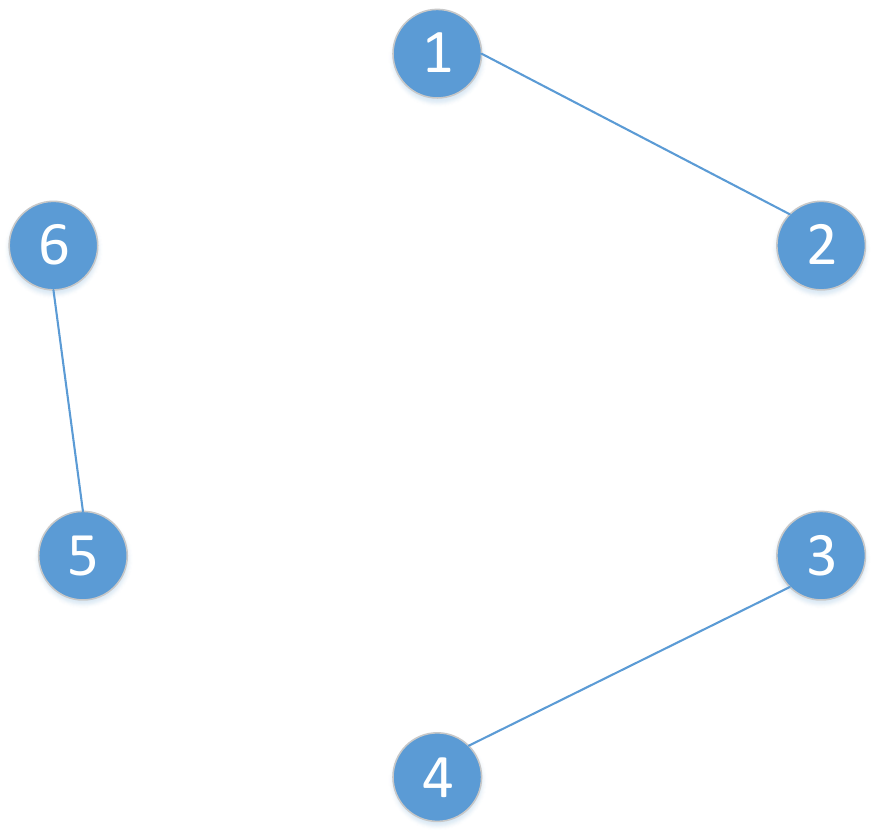}
				\caption{$\left\{\left\{1,2\right\},\left\{3,4\right\},\left\{5,6\right\}\right\}$}
    \end{subfigure}
    \begin{subfigure}[b]{0.3\textwidth}
        \includegraphics[scale=0.2]{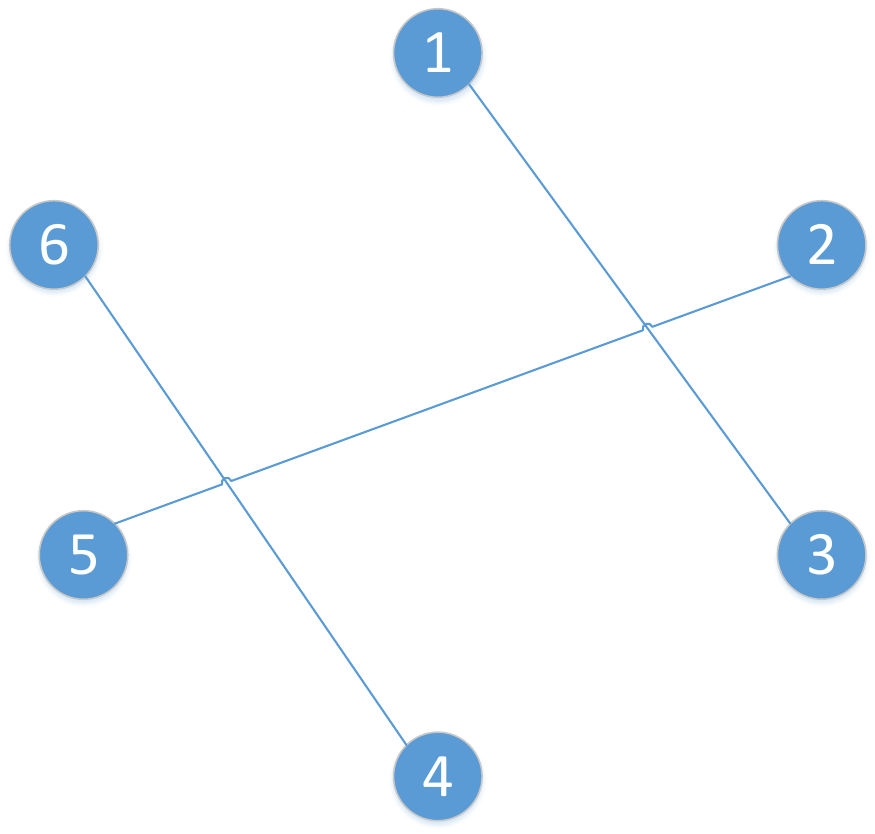}
				\caption{$\left\{\left\{1,3\right\},\left\{2,5\right\},\left\{4,6\right\}\right\}$}
        \end{subfigure}
    \begin{subfigure}[b]{0.3\textwidth}
        \includegraphics[scale=0.2]{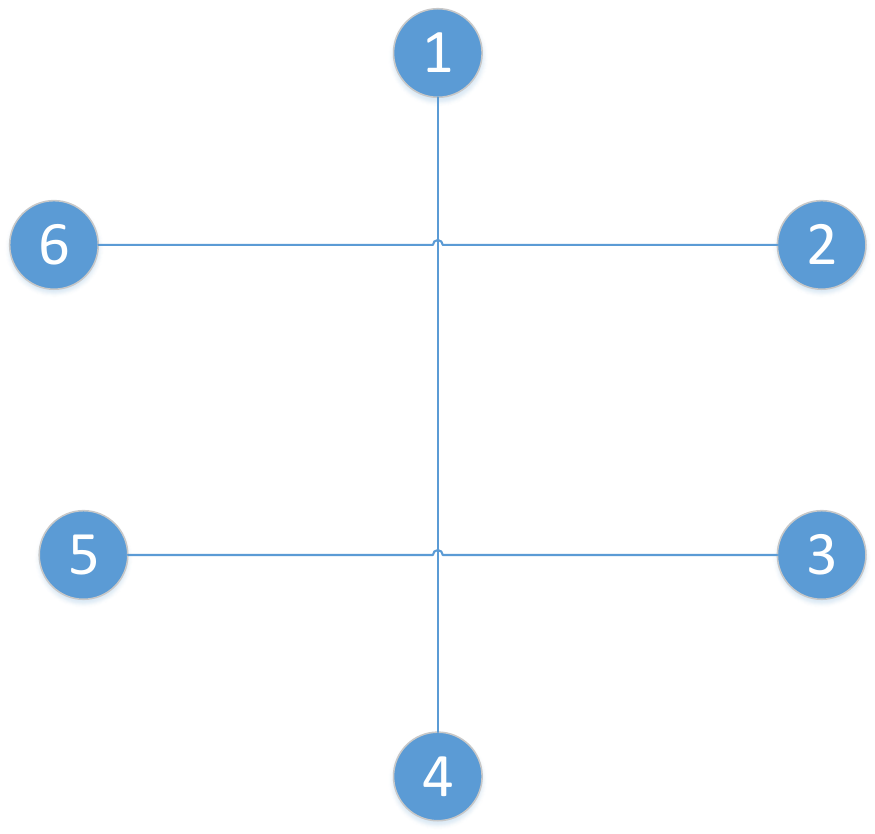}
				\caption{$\left\{\left\{1,4\right\},\left\{3,5\right\},\left\{2,6\right\}\right\}$}
    \end{subfigure}
		\begin{subfigure}[b]{0.3\textwidth}
        \includegraphics[scale=0.2]{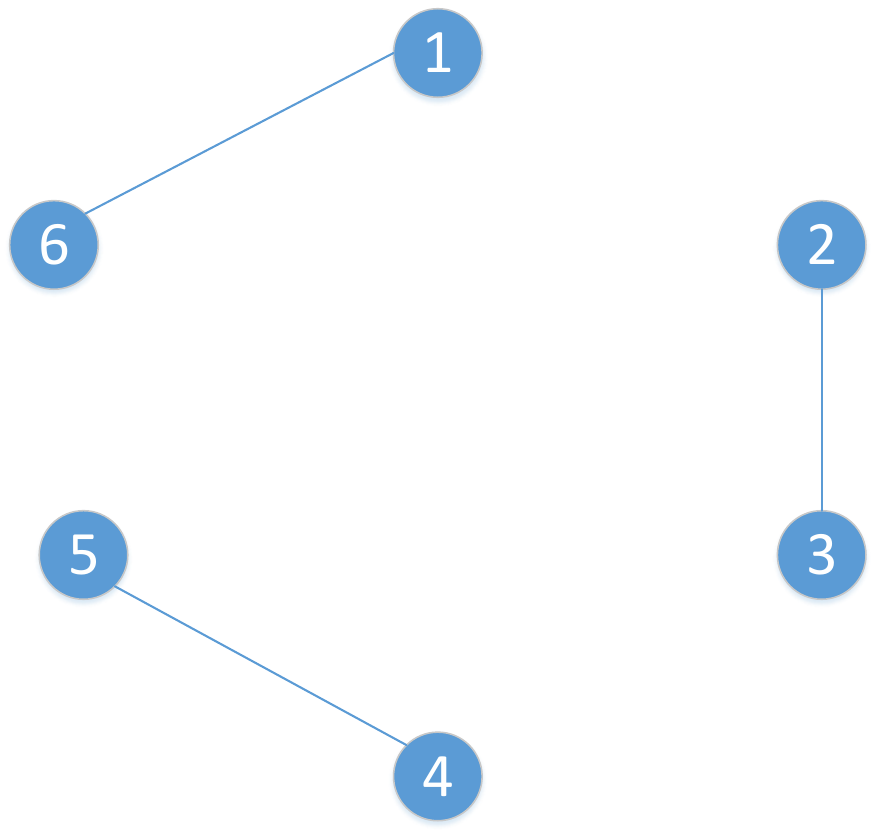}
				\caption{$\left\{\left\{1,6\right\},\left\{2,3\right\},\left\{4,5\right\}\right\}$}
    \end{subfigure}
	  \begin{subfigure}[b]{0.3\textwidth}
        \includegraphics[scale=0.2]{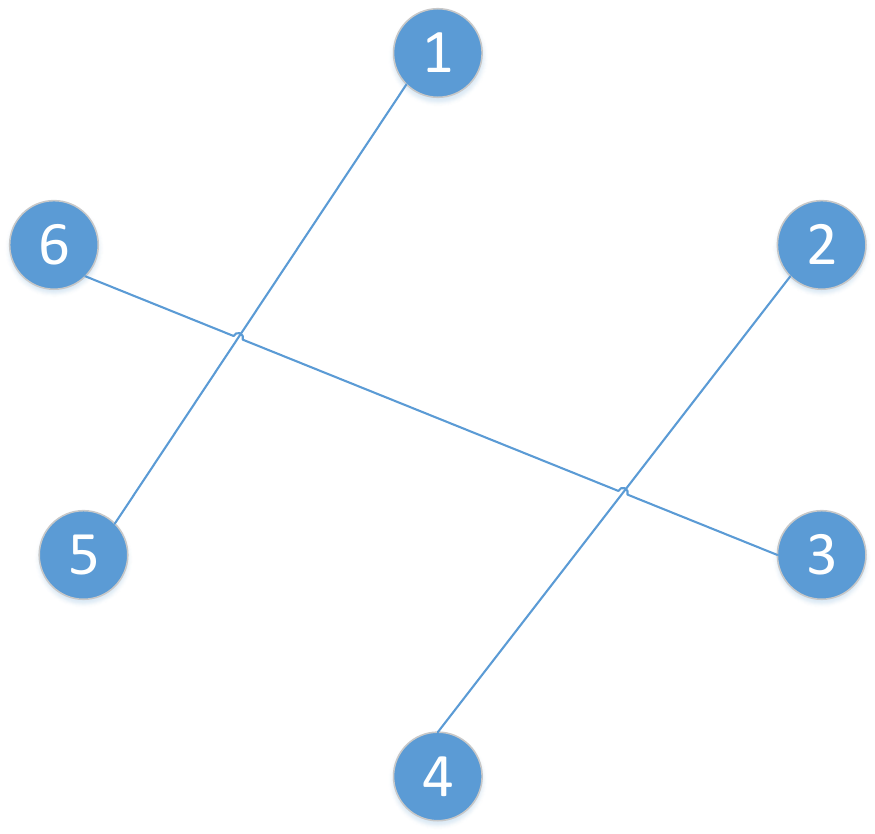}
				\caption{$\left\{\left\{1,5\right\},\left\{2,4\right\},\left\{3,6\right\}\right\}$}
    \end{subfigure}
    \caption{All perfect matchings and corresponding partitions for $\left\{1,2,3,4,5,6\right\}$}
		\label{partitioning}
\end{figure*}

\begin{definition}
A matching of a graph $G$ is perfect if it contains all of $G$'s vertices. From definition a perfect matching is a maximum matching although reverse is not necessarily true.
\end{definition}
\begin{definition}
1-factorization of a graph $G$ is a collection $\mathcal{D}$ of edge-disjoint perfect matchings (also referred to as 1-factors) whose union  is $\mathcal{E}(G)$. Graph $G$ is 1-factorable if it admits 1-factorization.
\end{definition}
\begin{lemma}
A complete graph $G$, with $V=2k$ for some  $k\in\mathbb{Z}^{+}$,  is 1-factorable and $\vert \mathcal{D}\vert=V$  
\end{lemma}
\begin{definition}
For a given edge $e_{i,j}$, $I(e_{i,j})$ is the 
index set of adjacent vertices $I(e_{i,j})=\left\{i,j\right\}$.
\end{definition}
\begin{definition}
A {\em partition} of a set $\mathcal{S}$ is a collection of nonempty and mutually disjoint subsets of $\mathcal{S}$, called \textit{blocks}, whose union is $\mathcal{S}$. For instance given set $\mathcal{S}=\left\{1,2,3,4\right\}$; $\left\{1\right\}$,$\left\{2,3,4\right\}$ is a partition of $\mathcal{S}$ with blocks of sizes $1$ and $3$.
\end{definition}
\begin{lemma}
Consider a graph $G$ with $V=2k$ for some  $k\in\mathbb{Z}^{+}$. Then, for any perfect matching $D$ of $G$, $\mathcal{P}\defeq\left\{I(e_{i,j}):e_{i,j}\in \mathcal{E}(D)\right\}$ is a partition of set $[V]$ with blocks of size two.
\end{lemma}
\begin{lemma}
Given set $[V]$, with $V=2k$ for some  $k\in\mathbb{Z}^{+}$, it is possible to obtain $V$ different disjoint partitions\footnote{By a disjoint partition, we mean that the distinct partitions, represented as sets, are disjoint.} of set $[V]$ with blocks of size two.
\end{lemma}
\begin{proof}
Consider a complete graph $G$, such that $\mathcal{V}(G)=[V]$ with $V=2k$. From Lemma 2, $\mathcal{E}(G)$ can be written as a collection $\mathcal{D}$ of edge-disjoint perfect matchings. Lemma 3 implies that for any $D\in\mathcal{D}$, one can obtain $\mathcal{P}\defeq\left\{I(e_{i,j}):e_{i,j}\in D\right\}$ which is a partition of set $[V]$ with  blocks of size two. Further, since any $D_{k},D_{l}\in \mathcal{D}$ edge-disjoint perfect matching, corresponding partitions, $\mathcal{P}_{k}\defeq\left\{I(e_{i,j}):e_{i,j}\in \mathcal{E}(D_{k})\right\}$ and $\mathcal{P}_{l}\defeq\left\{I(e_{i,j}):e_{i,j}\in \mathcal{E}(D_{l})\right\}$ are disjoint i.e., $\mathcal{P}_{k}\cap\mathcal{P}_{l}=\emptyset$.
\end{proof}
 In Fig. \ref{partitioning}, all disjoint partitions for set $[6]=\left\{1,2,3,4,5,6\right\}$, and the corresponding edge-disjoint perfect matchings are illustrated.  We note  that, although Lemma 4 is introduced for set $[V]$, with $V=2k$ for some  $k\in\mathbb{Z}^{+}$, it is valid for any set $\mathcal{K}$, $\vert\mathcal{K}\vert=V$, and we use the notation $\mathcal{P}^{\mathcal{K}}$  to refer a particular set $\mathcal{K}$.
 
 \begin{lemma}
Consider a complete graph $G$ with $V=2k+1$ for some  $k\in\mathbb{Z}^{+}$. Then, for any maximum matching $D$ of $G$,  $\mathcal{P}\defeq\left\{I(e_{i,j}):e_{i,j}\in \mathcal{E}(D)\right\}$ is a partition of set $[V]\setminus \left\{v\right\}$ with  blocks of size two, where $v\in \mathcal{V}(G) \setminus \mathcal{V}(D)$.
\end{lemma}

\begin{lemma}\label{lem:matching}
A complete graph $G$, with $V=2k+1$ for some  $k\in\mathbb{Z}^{+}$, can be decomposed into collection of $V$ edge-disjoint maximum matchings $\mathcal{D}$.    
\end{lemma}
\begin{figure}
    \centering
        \includegraphics[scale=0.35]{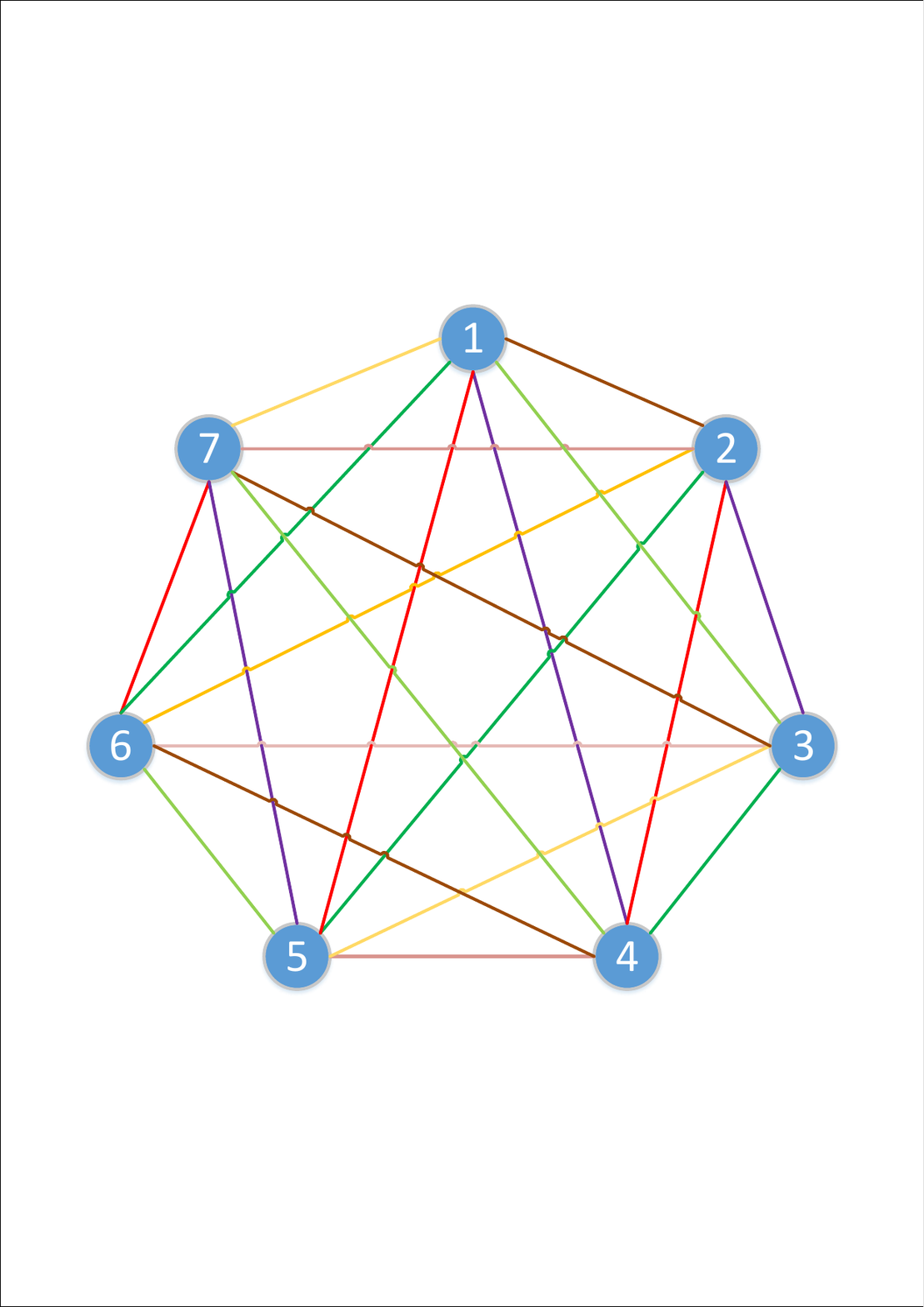}
				\caption{Each color represents a maximum matching.}
				\label{fig:maxmatch}
\end{figure}

\begin{proof}
Assume that all the vertices in $\mathcal{V}(G)$ are located on a circle according to their index with an increasing order. Then, for any vertex $j\in \mathcal{V}(G)$, a sub-graph $\tilde{G}$ with $\mathcal{V}(\tilde{G})=[V]\setminus\left\{j\right\}$ and $\mathcal{E}(\tilde{G})=\left\{e_{ \mod(j-1,V), \mod (j+1,V)},\ldots, e_{\mod (j-k ,V), \mod (j+k,V)}\right\}$, where $k=(V-1)/2$, is a maximum mathching. By following the aforementioned method, $\forall i\in \mathcal{V}(G)$, it is possible to generate a maximum matching $D_{i}$. Then one can easily observe  that for any $i,j$
\begin{equation}
\mathcal{E}(D_{i})\cap \mathcal{E}(D_{j}) =\emptyset,
\end{equation}
 and,
\begin{equation}
\cup_{i\in[V]} \mathcal{E}(D_{i}) = \mathcal{E}(G).
\end{equation}

\end{proof}
 Decomposition of a complete graph $G$, with $V=7$, into edge-disjoint maximum matchings is illustrated in  Figure \ref{fig:maxmatch}. From Lemma 5 and Lemma 6, we can conclude that for a given set $\mathcal{K}$, for each $i\in\mathcal{K}$, it is possible to have a partition of $\mathcal{K}\setminus\left\{i\right\}$ with blocks of size  two, denoted by $\mathcal{P}^{\mathcal{K}\setminus\left\{i\right\}}$ such that \begin{equation}
\mathcal{P}^{\mathcal{K}\setminus\left\{i\right\}}\cap\mathcal{P}^{\mathcal{K}\setminus\left\{j\right\}}= \emptyset \text{ for any } i,j\in\mathcal{K}. \nonumber
\end{equation}

\section{Construction procedure for $\Delta_{i,j}$ and $\mathcal{B}$ }\label{appendix_construc} 
In this section, we will explain the procedure for constructing the set of multicast messages, $\mathcal{B}$, that are sent in the last step of the delivery phase. The main concern of this procedure is to satisfy the constraint $\vert \Delta_{i,j} \vert= n_{i,j}$\footnote{Since the same partitioning is applied to all $\Lambda_{i}$'s, $\forall i\in \mathcal{K}^{h}$, $n_{i,j}=n_{j}$ for all $i\in \mathcal{K}^{h}$.}, $\forall i\in \mathcal{K}^{h}$ and $j\in \mathcal{K}^{l}$, while constructing the set $\mathcal{B}$. In the construction procedure, we consider two cases, where $k^{h}$ is even and odd, separately.
\begin{algorithm}[t]
    \SetKwInOut{Input}{Input}
    \SetKwInOut{Output}{Output}
    \Input{$\mathcal{K}^{h}$, $\left\{n_{j}\right\}_{j\in \mathcal{K}^{l}}$}
    \Output{$\left\{\Delta_{i,j}\right\}_{i\in \mathcal{K}^{h}, j\in \mathcal{K}^{l}}$, $\mathcal{B}$}
		$\Delta_{i,j}\gets\left\{\right\}$,  $\forall i\in \mathcal{K}^{h}, j\in \mathcal{K}^{l}$\;
		$\mathcal{B}\gets\left\{\right\}$\;
		\For{all $j\in \mathcal{K}^{l}$}{ 
		    construct $\mathcal{Q}^{n_{j}}$\\
                     \For{all $\left\{i,k\right\}\in \mathcal{Q}^{n_{j}}$}{
										  $\Delta_{i,j} \gets \Delta_{i,j}\bigcup\left\{W_{d_{i},\left\{k,j\right\}}\right\}$\;
											$\Delta_{k,j} \gets \Delta_{k,j}\bigcup\left\{W_{d_{k},\left\{i,j\right\}}\right\}$\;
											$\mathcal{B}\gets \mathcal{B}\bigcup \left\{W_{d_{i},\left\{k,j\right\}}\oplus W_{d_{k},\left\{i,j\right\}}\right\}$\;
										                                                      }
											   }
											\caption{ Set construction for the case of even $k^{h}$}
\label{alg:even_hlu}
\end{algorithm}
\subsection{Even number of high-level users}\label{ss:even}
Before presenting the algorithm for an even number of high-level users, we will briefly explain how the concept of {\em partitions} can be utilized for the construction of set $\mathcal{B}$. Consider $\mathcal{K}^{h}=\left\{1,2,3,4,5,6\right\}$ and a partition of  $\mathcal{K}^{h}$ with blocks of size two\footnote{In the scope of this work, we use the term partition to refer only a particular type of partition with blocks of size two. Hence, from this point on we use the term partition without further specifying the block sizes.}, e.g., $\mathcal{P}^{\mathcal{K}^{h}}=\left\{\left\{1,2\right\},\left\{3,4\right\},\left\{5,6\right\}\right\}$. Then, for a particular $j\in\mathcal{K}^{l}$, each $\left\{i,k\right\}\in\mathcal{P}^{\mathcal{K}^{h}}$ can be converted into the following multicast message\footnote{Throughout the paper, we often refer to the subset $\left\{i,k\right\}$ as a node pairing.}: $W_{d_{i},\left\{k,j\right\}}\oplus W_{d_{k},\left\{i,j\right\}}$. To this end, the specified partition $\mathcal{P}^{\mathcal{K}^{h}}$ corresponds to the following set of multicast messages;
\begin{equation}\label{exmppartition}
\left\{W_{d_{1},\left\{2,j\right\}}\oplus  W_{d_{2},\left\{1,j\right\}}, W_{d_{3},\left\{4,j\right\}}\oplus  W_{d_{4},\left\{3,j\right\}}, W_{d_{5},\left\{6,j\right\}}\oplus  W_{d_{6},\left\{5,j\right\}}\right\}.
\end{equation}
Recall that, $\mathcal{B}$ is the set of multicast-messages that are not decomposed and are sent in the last step of the delivery phase. Hence, if the multicast messages in (\ref{exmppartition}) are added to  $\mathcal{B}$, then sub-files $W_{d_{1},\left\{2,j\right\}}$,  $W_{d_{2},\left\{1,j\right\}}$, $W_{d_{3},\left\{4,j\right\}}$, $W_{d_{4},\left\{3,j\right\}}$,$W_{d_{5},\left\{6,j\right\}}$ and  $W_{d_{6},\left\{5,j\right\}}$ are added to sets $\Delta_{1,j}$, $\Delta_{2,j}$, $\Delta_{3,j}$, $\Delta_{4,j}$, $\Delta_{5,j}$, and $\Delta_{6,j}$, respectively. One can observe that, for a particular $j\in\mathcal{K}^{l}$, if a partition $\mathcal{P}^{\mathcal{K}^{h}}$ is used for adding multicast messages to set $\mathcal{B}$, then exactly one sub-file is added to the set $\Delta_{i,j}$ for each  $i \in \mathcal{K}^{h}$. Hence, for a particular $j\in\mathcal{K}^{l}$, we consider any $n_{j}$ disjoint partitions $\mathcal{P}^{\mathcal{K}^{h}}_{1},\ldots,\mathcal{P}^{\mathcal{K}^{h}}_{n_{j}}$ to determine the multicast messages to be included in set $\mathcal{B}$, and the constraint $\vert\Delta_{i,j}\vert=n_{j}$ would be satisfied for all $i \in \mathcal{K}^{h}$. Hence, we define $\mathcal{Q}^{n_{j}}= \cup_{i=1 : n_{j}}\mathcal{P}^{\mathcal{K}^{h}}_{i}$, and use it to add  multicast messages to set $\mathcal{B}$ in Algorithm \ref{alg:even_hlu}. We note that $\mathcal{Q}^{n_{j}}$ can be considered as a set of user pairings, where each user appears in exactly $n_{j}$ pairings. Further details on the construction of the disjoint partitions  $\mathcal{P}^{\mathcal{K}^{h}}_{1},\ldots,\mathcal{P}^{\mathcal{K}^{h}}_{n}$ are explained in Appendix \ref{appendix_partition}.
\subsection{Odd number of high-level users}\label{ss:odd}
\begin{algorithm}[t]
    \SetKwInOut{Input}{Input}
    \SetKwInOut{Output}{Output}
    \Input{$\mathcal{K}^{h},n_{even}$}
    \Output{$\mathcal{Q}^{n_{even}}$}
    $\tilde{\mathcal{K}}^{h}\subset\mathcal{K}^{h}$ with $\vert \tilde{\mathcal{K}}^{h}\vert = k^{h}-1$\;
		\For{$j=1:n_{even}/2$}{
		    $\left\{i,k\right\} \gets \mathcal{P}^{\tilde{\mathcal{K}}^{h}}(j)$\;
				$\mathcal{Q}^{n_{even}} \gets \mathcal{Q}^{n_{even}}\bigcup (\mathcal{P}^{\mathcal{K}^{h}\setminus\left\{i\right\}}\bigcup \mathcal{P}^{\mathcal{K}^{h}\setminus\left\{k\right\}}\bigcup\left\{i,k\right\})$\;
											   }
	 										\caption{ Construction of $\mathcal{Q}^{n_{even}}$ for odd $k^{h}$}
\label{alg:n_even}
\end{algorithm}
Above, we first obtained a set of node pairings $\mathcal{Q}^{n_{j}}$ for each $j\in\mathcal{K}^{l}$, and then constructed the set of multicast messages $\mathcal{B}$ using these node pairings in Algorithm \ref{alg:even_hlu}. We recall that $\mathcal{Q}^{n_{j}}$ is the union of $n_{j}$ partitions generated from $\mathcal{K}^{h}$. However, when $k^{h}$ is odd,  it is not possible to obtain partition of $\mathcal{K}^{h}$ with blocks of size two. We remark that, if $k^{h}$ is  odd, $k^{l}$ must be even, which means that for $k^{l}/2$ low-level users, $n_{j}$ will be an odd number $n_{odd}$, and for the remaining low-level users, $n_{j}$ will be an even number $n_{even}$. Let $\mathcal{K}^{l}_{odd}$ and $\mathcal{K}^{l}_{even}$ be the subset of low-level users with $n_{odd}$ and $n_{even}$, respectively. Furthermore, let  $k^{l}_{odd}$ and $k^{l}_{even}$ denote the cardinality of the sets $\mathcal{K}^{l}_{odd}$ and $\mathcal{K}^{l}_{even}$, respectively.\\
\indent For the case of $n_{even}$, we introduce a new method, Algorithm \ref{alg:n_even}, to construct $\mathcal{Q}^{n_{even}}$ for each $j\in\mathcal{K}^{l}$. In Algorithm \ref{alg:n_even}, we use the notation $\mathcal{A}(i)$ to denote the $i$th element of set $\mathcal{A}$ for an arbitrary ordering of its elements\footnote{ Although we use index for the set $\mathcal{P}^{\tilde{\mathcal{K}}^{h}}$, Algorithm \ref{alg:n_even} does not require a particular ordering for this set.}. We also want to remark that the partitions  used in Algorithm \ref{alg:n_even} are disjoint, i.e., 
\begin{equation}
\mathcal{P}^{\mathcal{K}^{h}\setminus\left\{i\right\}}\cap\mathcal{P}^{\mathcal{K}^{h}\setminus\left\{k\right\}}=\emptyset,
\end{equation}
where  $i,k \in \mathcal{K}^{h}$, and $i\neq k$. The process of obtaining these disjoint partitions are explained in Appendix \ref{appendix_partition}. Now, one can easily observe that, each high-level user appears in exactly one pairing in $\mathcal{P}^{\mathcal{K}^{h}\setminus\left\{i\right\}}$,  except $i$, and in $\mathcal{P}^{\mathcal{K}^{h}\setminus\left\{i\right\}}\bigcup \mathcal{P}^{\mathcal{K}^{h}\setminus\left\{k\right\}}\cup\left\{i,k\right\}$  each high-level user appears exactly in two pairings. To clarify, assume that we want to construct $\mathcal{Q}^{4}$ for $\mathcal{K}^{h}=\left\{1,2,3,4,5,6,7\right\}$. We can set $\tilde{\mathcal{K}}^{h}=\left\{1,2,3,4,5,6\right\}$ and use partition  $\mathcal{P}^{\tilde{\mathcal{K}}^{h}}=\left\{\left\{1,6\right\},\left\{2,5\right\},\left\{3,4\right\}\right\}$, so that $\mathcal{P}^{\tilde{\mathcal{K}}^{h}}(1)=\left\{1,6\right\}$ and $\mathcal{P}^{\tilde{\mathcal{K}}^{h}}(2)=\left\{2,5\right\}$. Then, Algorithm \ref{alg:n_even} uses the partitions given in Table \ref{partitionsexample} to construct $\mathcal{Q}^{4}$ as follows:
\begin{equation}
\begin{aligned}
\mathcal{Q}^{4}=&\left\{ \left\{2,7\right\},\left\{3,6\right\},\left\{4,5\right\},\left\{2,3\right\},\left\{1,4\right\},\left\{5,7\right\},\left\{1,6\right\}\right. \\
&\left. \left\{1,3\right\},\left\{4,7\right\},\left\{5,6\right\},\left\{4,6\right\},\left\{3,7\right\},\left\{1,2\right\},\left\{2,5\right\}\right\}.
\end{aligned}
\end{equation}

\begin{table}{\footnotesize
    \begin{center}
    \begin{tabular}{ | l | p{3cm} |}
    \hline
    Partition & Corresponding set \\ \hline
	$\mathcal{P}^{\mathcal{K}^{h}\setminus\left\{1\right\}}$ & $\left\{\left\{2,7\right\},\left\{3,6\right\},\left\{4,5\right\}\right\}$ \\ \hline
    $\mathcal{P}^{\mathcal{K}^{h}\setminus\left\{6\right\}}$ & $\left\{\left\{2,3\right\},\left\{1,4\right\},\left\{5,7\right\}\right\}$  \\ \hline
     $\mathcal{P}^{\mathcal{K}^{h}\setminus\left\{2\right\}}$ & $\left\{\left\{1,3\right\},\left\{4,7\right\},\left\{5,6\right\}\right\}$\\ \hline
      $\mathcal{P}^{\mathcal{K}^{h}\setminus\left\{5\right\}}$ & $\left\{\left\{4,6\right\},\left\{3,7\right\},\left\{1,2\right\}\right\}$\\ \hline
		\end{tabular}
		\caption{Partitions used to construct $\mathcal{Q}^{4}$ for $\mathcal{K}^{h}=\left\{1,2,3,4,5,6,7\right\}$.}
		\label{partitionsexample}	          
		\end{center}
		       }
\end{table}

Eventually, in $\mathcal{Q}^{n_{even}}$ each high-level user appears exactly in $n_{even}$ pairings. Hence, as in Algorithm \ref{alg:even_hlu}, $\mathcal{Q}^{n_{even}}$ can be used to construct sets $\Delta_{i,j}$ as well the set of multicast messages $\mathcal{B}$.\\  
\indent We remark that, for each $j\in \mathcal{K}^{l}_{even}$ the same set of node pairings $\mathcal{Q}^{n_{even}}$ is used, thus the process is identical for each $j\in \mathcal{K}^{l}_{even}$. However, for low-level users $j\in\mathcal{K}^{l}_{odd}$ the process will not be identical since it is not possible to construct a single $\mathcal{Q}^{n_{odd}}$ for all $j\in\mathcal{K}^{l}_{odd}$. Nevertheless, we follow a similar procedure to construct the multicast messages for the low-level users in $\mathcal{K}^{l}_{odd}$.\\ 
\indent The overall procedure for the case of odd number of high-level users is illustrated in Algorithm \ref{alg:odd_kh}. In the algorithm, the set of node pairings $\mathcal{Q}_{j}$ for each $j\in\mathcal{K}^{l}_{odd}$ is constructed separately, and is used to decide the multicast message to be placed in $\mathcal{B}$, and the sub-files to be placed in sets $\Delta_{i,j}$, as in  Algorithm 1. From the construction, one can easily observe that in $\mathcal{Q}_{j}$, each high-level user appears exactly in $n_{odd}$ pairings except a particular $k$ that appears in $n_{odd}-1$ pairings. Thus, when $\mathcal{Q}_{j}$ is used to construct multicast messages, constraint $\vert\Delta_{i,j}\vert=n_{odd}$ is satisfied for all $i\in\mathcal{K}^{h}$, except $i=k$. Therefore, at line 11 and 16 in Algorithm \ref{alg:odd_kh}, we add a sub-file to set $\Delta_{k,j}$, where the high-level user $k$  appears in $n_{odd}-1$ pairings in $\mathcal{Q}_{j}$, and at line 17 we XOR these sub-files. Hence, eventually we ensure that, at the end of the algorithm the equality constraint $\vert \Delta_{i,j} \vert=n_{odd}$ is satisfied for all $i\in\mathcal{K}^{h}$ and for all $j\in \mathcal{K}^{l}_{odd}$. \\
\indent With the proposed set construction algorithms we are ensuring that, in the third and fourth steps of the delivery phase all the sub-files are delivered with a multicasting gain of two\footnote{When $\vert\cup_{i\in\mathcal{K}^{h},j\in\mathcal{K}^{l} }\Delta_{i,j}\vert$ is odd, exactly one sub-file is unicasted, while the remaining sub-files achieve a multicasting gain of two, as in Example 1.}, which explains the achievable delivery rate.\\ 
 \begin{algorithm}[t]
    \SetKwInOut{Input}{Input}
    \SetKwInOut{Output}{Output}
    \Input{$\mathcal{K}^{l}_{odd},\mathcal{K}^{l}_{even},\mathcal{K}^{h}$, $\left\{n_{even},n_{odd}\right\}$}
    \Output{$\mathcal{B},\left\{\Delta_{i,j}\right\}_{i\in K^{h}, j\in K^{l}}$}
		$\Delta_{i,j}\gets\left\{\right\}$ $\forall$ $i\in K^{h}, j\in K^{l}$, $\mathcal{Q}_{j} \gets \left\{\right\}$ $\forall j\in K^{l}$\;
        $\tilde{\mathcal{K}}^{h}\subset\mathcal{K}^{h}$ with  $\vert \tilde{\mathcal{K}}^{h}\vert = k^{h}-1$\; 
		construct $\mathcal{Q}^{n_{odd}-1}$ and construct $\mathcal{Q}^{n_{even}}$ 	\;
		$\mathcal{Q}_{j} \gets \mathcal{Q}^{n_{even}}$,  $\forall j\in \mathcal{K}^{l}_{even}$\;
        $\mathcal{Q}_{j} \gets \mathcal{Q}^{n_{odd}-1}$,  $\forall j\in \mathcal{K}^{l}_{odd}$\;
        $\tilde{n} \gets \left\lceil n_{odd}/2\right\rceil$\;
		\For{$m=1:k^{l}_{odd}$} {		            
                \If{ $m$ is odd}{
								 $\left\{i,k\right\} \gets \mathcal{P}^{\tilde{\mathcal{K}}^{h}}(\tilde{n})$\;
								 $\mathcal{Q}_{\mathcal{K}^{l}_{odd}(m)} \gets \mathcal{Q}_{\mathcal{K}^{l}_{odd}(m)}\bigcup \mathcal{P}^{\mathcal{K}^{h}\setminus\left\{i\right\}}$\;
								 $\Delta_{i,\mathcal{K}^{l}_{odd}(m)} \gets \Delta_{i,\mathcal{K}^{l}_{odd}(m)}\bigcup\left\{W_{d_{i},\left\{k,\mathcal{K}^{l}_{odd}(m)\right\}}\right\}$\;											
                                } 
							  \If{ $m$ is even}{
								 $\left\{i,k\right\} \gets \mathcal{P}^{\tilde{\mathcal{K}}^{h}}(\tilde{n})$\;
								 $\mathcal{Q}_{\mathcal{K}^{l}_{odd}(m)} \gets \mathcal{Q}_{\mathcal{K}^{l}_{odd}(m)}\bigcup \mathcal{P}^{\mathcal{K}^{h}\setminus\left\{k\right\}}$\;
								 $\Delta_{k,\mathcal{K}^{l}_{odd}(m)} \gets \Delta_{k,\mathcal{K}^{l}_{odd}(m)}\bigcup\left\{W_{d_{k},\left\{i,\mathcal{K}^{l}_{odd}(m)\right\}}\right\}$\;
								 $\mathcal{B}\gets \mathcal{B}\bigcup \left\{W_{d_{i},\left\{k,\mathcal{K}^{l}_{odd}(m-1)\right\}}\oplus W_{d_{k},\left\{i,\mathcal{K}^{l}_{odd}(m)\right\}}\right\}$\;
                                }

					                  }
		\For{all $j\in K^{l}$}{
		                      \For{all $(i,k)\in \mathcal{Q}_{j}$}{
		                        $\Delta_{i,j} \gets \Delta_{i,j}\bigcup\left\{W_{d_{i},\left\{k,j\right\}}\right\}$\;
				                    $\Delta_{k,j} \gets \Delta_{k,j}\bigcup\left\{W_{d_{k},\left\{i,j\right\}}\right\}$\;
				                    $\mathcal{B}\gets \mathcal{B}\bigcup \left\{W_{d_{i},\left\{k,j\right\}}\oplus W_{d_{k},\left\{i,j\right\}}\right\}$\;
										                                                 }											
													}
														
												\caption{ Set construction for the case of odd $k^{h}$}
												\label{alg:odd_kh}
										
\end{algorithm}
\indent Now, we go back to Example 1, where we have an odd number of high-level files. In particular, we have $\mathcal{K}^{h}=\left\{ 1,2,3,4,5 \right\}$, $\mathcal{K}^{l}_{odd}=\left\{6\right\}$ and $\mathcal{K}^{l}_{even}=\left\{7\right\}$, thus $n_{odd}=1$ and $n_{even}=2$. Further, partition are given as $\mathcal{P}^{\mathcal{K}^{h}\setminus\left\{1\right\}}=\left\{\left\{2,5\right\},\left\{3,4\right\}\right\}$, $\mathcal{P}^{\mathcal{K}^{h}\setminus\left\{2\right\}}=\left\{\left\{1,3\right\},\left\{4,5\right\}\right\}$, $\mathcal{P}^{\mathcal{K}^{h}\setminus\left\{3\right\}}=\left\{\left\{1,5\right\},\left\{2,4\right\}\right\}$, $\mathcal{P}^{\mathcal{K}^{h}\setminus\left\{4\right\}}=\left\{\left\{3,5\right\},\left\{1,2\right\}\right\}$, $\mathcal{P}^{\mathcal{K}^{h}\setminus\left\{5\right\}}=\left\{\left\{1,4\right\},\left\{3,2\right\}\right\}$. Then, Algorithm \ref{alg:odd_kh} takes the set $\tilde{\mathcal{K}}^{h}=\left\{ 1,2,3,4\right\}$ and use the partition  $\mathcal{P}^{\mathcal{K}^{h}}=\mathcal{P}^{\mathcal{K}^{h}\setminus\left\{5\right\}}=\left\{\left\{1,4\right\},\left\{2,3\right\}\right\}$. Accordingly, Algorithm \ref{alg:odd_kh} constructs $\mathcal{Q}^{n_{odd}-1}=\emptyset$ and $\mathcal{Q}^{n_{even}}=\left\{\left\{1,2\right\},\left\{1,4\right\},\left\{2,5\right\},\left\{3,4\right\},\left\{3,5\right\} \right\}$. Thereafter, sets $\mathcal{Q}_{6}$ and $\mathcal{Q}_{7}$ are constructed as  $\mathcal{Q}_{6}=\left\{\left\{2,5\right\},\left\{3,4\right\}\right\}$ and $\mathcal{Q}_{7}=\left\{\left\{1,2\right\},\left\{1,4\right\},\left\{2,5\right\},\left\{3,4\right\},\left\{3,5\right\} \right\}$. The corresponding set of multicast messages, $\mathcal{B}$, are already illustrated with blue in Table \ref{Conv_delivery}.

\section{Partition of $\Lambda$}\label{approx_part}
\begin{figure}
    \centering
        \includegraphics[scale=0.25]{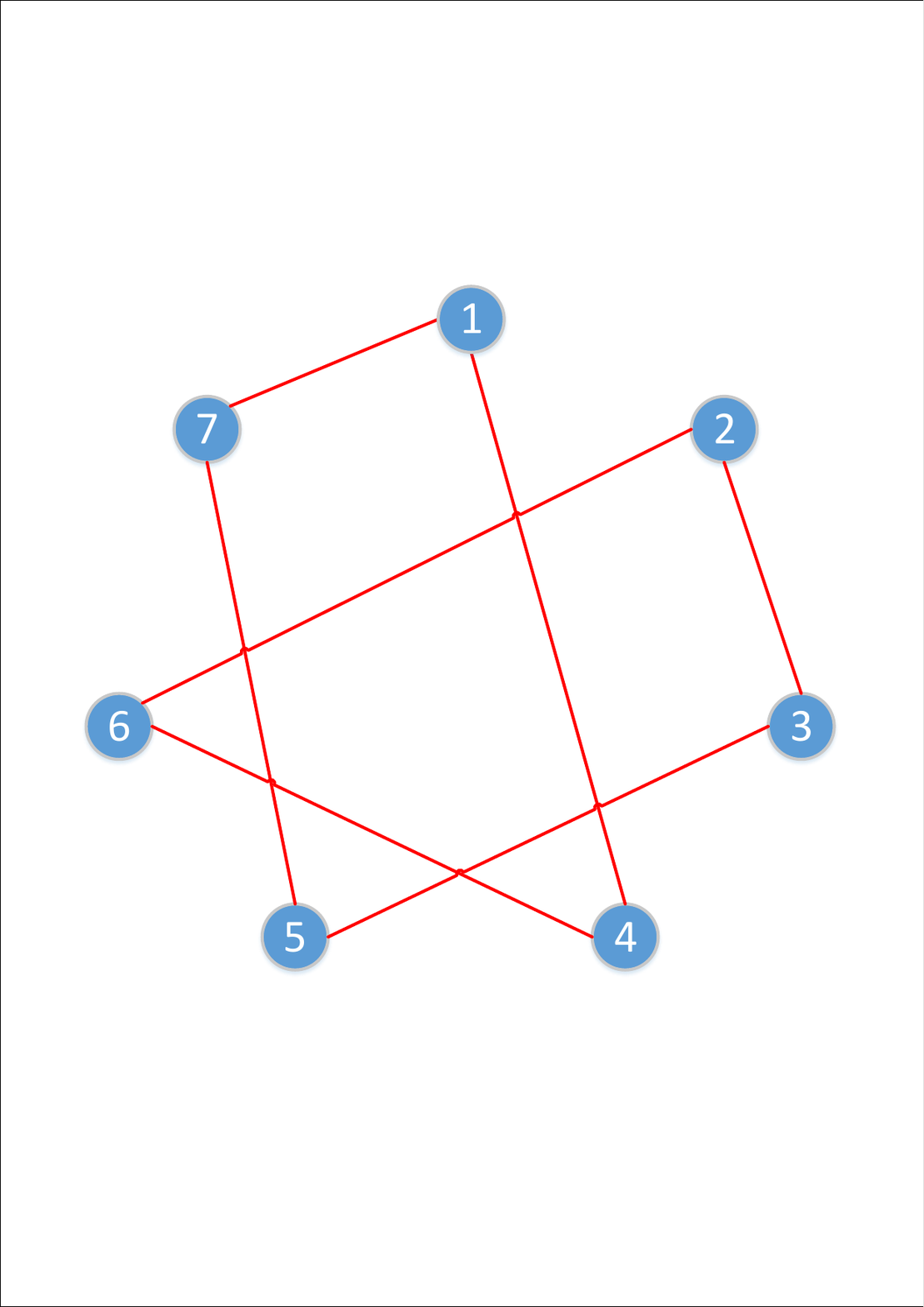}
				\caption{Hamiltonian cycle corresponding to $\mathcal{Q}=\left\{\left\{1,7\right\},\left\{7,5\right\},\left\{2,6\right\},\left\{2,3\right\},\left\{1,4\right\},\left\{6,4\right\},\left\{3,5\right\}\right\}$. }
				\label{hamiltonian}
\end{figure}

 In this part, we will show how $\Lambda_{i}=\left\{W_{d_{i},\left\{k,j\right\}}:k,j\in \mathcal{K}^{l}\right\}$ can be approximately partitioned. Recall that the main concern is to assign sub-file $W_{d_{i},\left\{k,j\right\}}$ to either $\Lambda_{i,k}$ or $\Lambda_{i,j}$ in order to achieve approximately uniform cardinality  sets  $\left\{\Lambda_{i,j}\right\}_{j\in \mathcal{K}^{l}}$. We consider two cases, where $k^{l}$ is  even and odd, respectively. We start with the case where $k^{l}$ is an odd number. Let the set of partitions $\left\{\mathcal{P}^{\mathcal{K}^{l}\setminus\left\{i\right\}}\right\}_{i\in \mathcal{K}^{l}}$ are given for $\mathcal{K}^{l}$. Then Algorithm \ref{alg:partition_lambda} is used to construct $\left\{\Lambda_{i,j}\right\}_{j\in \mathcal{K}^{l}}$.
\begin{algorithm}
    \SetKwInOut{Input}{Input}
    \SetKwInOut{Output}{Output}
    \Input{$\left\{\mathcal{P}^{\mathcal{K}^{l}\setminus\left\{j\right\}}\right\}_{j\in \mathcal{K}^{l}}$,$\Lambda_{i}$}
    \Output{$\left\{\Lambda_{i,j}\right\}_{j\in K^{l}}$}
    Pick an element $s\in \mathcal{K}^{l}$ randomly\\
		\For{$m=1:(k^{l}-1)/2$}{
		    $\left\{j,k\right\} \gets \mathcal{P}^{\mathcal{K}^{l}\setminus\left\{s\right\}} (m)$\;
				$\mathcal{Q} \gets \mathcal{P}^{\mathcal{K}^{l}\setminus\left\{j\right\}}\bigcup \mathcal{P}^{\mathcal{K}^{l}\setminus\left\{k\right\}} \bigcup \left\{j,k\right\}$\;
				$\mathcal{W} \gets \left\{W_{d_{i},\left\{k,j\right\}}: \right\{k,j\left\}\in \mathcal{Q}\right\}$\;
				Distribute $W$ to sets $\left\{\Lambda_{i,j}\right\}_{j\in K^{l}}$ sequentially 
											   }
											\caption{Partition of $\Lambda_{i}$}
                                            \label{alg:partition_lambda}
\end{algorithm}
Note that, in each step of Algorithm \ref{alg:partition_lambda} the size of set $\mathcal{W}$ is equal to $k^{l}$ and  low-level user index $j$ appears in exactly two sub-files in the $\mathcal{W}$. By "sequential distribution'' we mean that we start with some low-level user $j$ and  take the two sub-files $W_{d_{i},\left\{k,j\right\}},W_{d_{i},\left\{\acute{k},j\right\}}\in\mathcal{W}$. We start with assigning $W_{d_{i},\left\{k,j\right\}}$ to set $\Lambda_{i,j}$, then the other sub-file $W_{d_{i},\left\{\acute{k},j\right\}}$ is assigned to set $\Lambda_{i,\acute{k}}$. Thereafter, we take the file $W_{d_{i},\left\{\acute{k},\acute{j}\right\}}$ and assign it to set $\Lambda_{i,\acute{j}}$ and we continue the process in the same way. Formally speaking,  each $\mathcal{Q}$ in the algorithm resembles a Hamiltonian cycle, and in Algorithm \ref{alg:partition_lambda} we are assigning each edge in the cycle to a node. To clarify, consider Hamiltonian cycle corresponding to $\mathcal{Q}=\left\{\left\{1,7\right\},\left\{7,5\right\},\left\{2,6\right\},\left\{2,3\right\},\left\{1,4\right\},\left\{6,4\right\},\left\{3,5\right\}\right\}$ is illustrated in  Fig. \ref{hamiltonian}. Assume that, we start from node 1 and assign $e_{1,7}$ to node 1, and edges $e_{7,5},e_{3,5},e_{2,3},e_{2,6},e_{6,4}$,and $e_{1,4}$ are assigned to nodes $7,5,3,2,6,4$, respectively. Note that,  each edge corresponds to a sub-file and each node corresponds to a set $\Lambda_{i,j}$. 
We further remark that a complete graph $G$ with odd order can be decomposed into Hamiltonian cycles (as done in  Algorithm \ref{alg:partition_lambda}); hence, $\Lambda_{i}$ can be partitioned uniformly where $\vert\Lambda_{i,j}\vert=(k_{l}-1)/2,~ \forall i\in\mathcal{K}^{h},~\forall j\in\mathcal{K}^{h}.$\\
\indent For the case of even $k^{l}$ we can use a similar approach, but this time we construct the Hamiltonian cycles via combining two perfect matchings. Different from the previous case complete graph G with even order cannot be decomposed to Hamiltonian cycles. Hence, in the end we will have a remaining perfect matching. Those edges in the last perfect matching are assigned to nodes randomly. Therefore, in the end, $k^{l}/2$ of the partitions will have cardinality $k^{l}/2-1$, while the remaining will have cardinality $k^{l}/2$.
\bibliographystyle{IEEEtran}
\bibliography{IEEEabrv,ref}

\begin{thebibliography}{10}
\providecommand{\url}[1]{#1}
\csname url@samestyle\endcsname
\providecommand{\newblock}{\relax}
\providecommand{\bibinfo}[2]{#2}
\providecommand{\BIBentrySTDinterwordspacing}{\spaceskip=0pt\relax}
\providecommand{\BIBentryALTinterwordstretchfactor}{4}
\providecommand{\BIBentryALTinterwordspacing}{\spaceskip=\fontdimen2\font plus
\BIBentryALTinterwordstretchfactor\fontdimen3\font minus
  \fontdimen4\font\relax}
\providecommand{\BIBforeignlanguage}[2]{{%
\expandafter\ifx\csname l@#1\endcsname\relax
\typeout{** WARNING: IEEEtran.bst: No hyphenation pattern has been}%
\typeout{** loaded for the language `#1'. Using the pattern for}%
\typeout{** the default language instead.}%
\else
\language=\csname l@#1\endcsname
\fi
#2}}
\providecommand{\BIBdecl}{\relax}
\BIBdecl

\bibitem{CD.F1}
M.~A. Maddah-Ali and U.~Niesen, ``Fundamental limits of caching,'' \emph{IEEE
  Trans. Inf. Theory}, vol.~60, no.~5, May 2014.

\bibitem{CD.F2}
------, ``Decentralized coded caching attains order-optimal memory-rate
  tradeoff,'' \emph{IEEE/ACM Trans. Netw.}, vol.~23, no.~4, Aug 2015.

\bibitem{CD.F3}
Q.~Yu, M.~A. Maddah-Ali, and A.~S. Avestimehr, ``The exact rate-memory tradeoff
  for caching with uncoded prefetching,'' \emph{IEEE Transactions on
  Information Theory}, vol.~64, no.~2, pp. 1281--1296, Feb 2018.

\bibitem{CD.NCS1}
M.~M. Amiri, Q.~Yang, and D.~G{\"u}nd{\"u}z, ``Decentralized caching and coded
  delivery with distinct cache capacities,'' \emph{IEEE Transactions on
  Communications}, vol.~65, no.~11, pp. 4657--4669, Nov 2017.

\bibitem{CD.NCS2}
A.~M. {Ibrahim}, A.~A. {Zewail}, and A.~{Yener}, ``Coded caching for
  heterogeneous systems: An optimization perspective,'' \emph{IEEE Transactions
  on Communications}, vol.~67, no.~8, pp. 5321--5335, 2019.

\bibitem{CD.NCS3}
------, ``Device-to-device coded-caching with distinct cache sizes,''
  \emph{IEEE Transactions on Communications}, pp. 1--1, 2020.

\bibitem{CD.ULR1}
A.~Tang, S.~Roy, and X.~Wang, ``Coded caching for wireless backhaul networks
  with unequal link rates,'' \emph{IEEE Transactions on Communications},
  vol.~PP, no.~99, pp. 1--1, 2017.

\bibitem{CD.ULR3}
M.~M. Amiri and D.~G{\"u}nd{\"u}z, ``Cache-aided content delivery over erasure
  broadcast channels,'' \emph{IEEE Transactions on Communications}, vol.~PP,
  no.~99, pp. 1--1, 2017.

\bibitem{CD.ND1}
U.~Niesen and M.~A. Maddah-Ali, ``Coded caching with nonuniform demands,''
  \emph{IEEE Trans. Inf. Theory}, vol.~63, no.~2, Feb 2017.

\bibitem{CD.ND2}
M.~Ji, A.~M. Tulino, J.~Llorca, and G.~Caire, ``Order-optimal rate of caching
  and coded multicasting with random demands,'' \emph{IEEE Trans. Inf. Theory},
  vol.~63, no.~6, June 2017.

\bibitem{CD.ND3}
J.~Zhang, X.~Lin, and X.~Wang, ``Coded caching under arbitrary popularity
  distributions,'' \emph{IEEE Transactions on Information Theory}, vol.~64,
  no.~1, pp. 349--366, Jan 2018.

\bibitem{CD.ND4}
J.Hachem, N.~Karamchandani, and S.~N. Diggavi, ``Coded caching for multi-level
  popularity and access,'' \emph{IEEE Trans. Inf. Theory}, vol.~63, no.~5, May
  2017.

\bibitem{CD.ND5}
A.~M. {Daniel} and W.~{Yu}, ``Optimization of heterogeneous coded caching,''
  \emph{IEEE Transactions on Information Theory}, vol.~66, no.~3, pp.
  1893--1919, 2020.

\bibitem{CD.ND6}
S.~{Jin}, Y.~{Cui}, H.~{Liu}, and G.~{Caire}, ``Structural properties of
  uncoded placement optimization for coded delivery,'' \emph{CoRR}, vol.
  abs/1707.07146, 2017.

\bibitem{CD.ND7}
A.~Ramakrishnan, C.~Westphal, and A.~Markopoulou, ``An efficient delivery
  scheme for coded caching,'' in \emph{Proceedings of the 2015 27th
  International Teletraffic Congress}, ser. ITC '15.\hskip 1em plus 0.5em minus
  0.4em\relax Washington, DC, USA: IEEE Computer Society, 2015, pp. 46--54.

\bibitem{VS1}
X.~Cheng, C.~Dale, and J.~Liu, ``Statistics and social network of youtube
  videos,'' in \emph{2008 16th Int. Workshop on Quality of Service}, June 2008.

\bibitem{VS2}
M.~Cha, H.~Kwak, P.~Rodriguez, Y.~Y. Ahn, and S.~Moon, ``Analyzing the video
  popularity characteristics of large-scale user generated content systems,''
  \emph{IEEE/ACM Trans. Netw.}, vol.~17, no.~5, Oct 2009.

\bibitem{VS3}
M.~Zink, K.~Suh, Y.~Gu, and J.~Kurose, ``Characteristics of youtube network
  traffic at a campus network - measurements, models, and implications,''
  \emph{Comput. Netw.}, vol.~53, no.~4, pp. 501--514, Mar. 2009.

\bibitem{VS4}
J.~Lin, Z.~Li, G.~Xie, Y.~Sun, K.~Salamatian, and W.~Wang, ``Mobile video
  popularity distributions and the potential of peer-assisted video delivery,''
  \emph{IEEE Communications Magazine}, vol.~51, no.~11, pp. 120--126, November
  2013.

\bibitem{CD.Dopt1}
K.~Wan, D.~Tuninetti, and P.~Piantanida, ``Novel delivery schemes for
  decentralized coded caching in the finite file size regime,'' in \emph{2017
  IEEE International Conference on Communications Workshops (ICC Workshops)},
  May 2017, pp. 1183--1188.

\bibitem{CD.Dopt2}
N.~Zhang and M.~Tao, ``Fitness-aware coded multicasting for decentralized
  caching with finite file packetization,'' \emph{IEEE Wireless Communications
  Letters}, pp. 1--1, 2018.

\bibitem{DCU}
M.~S. {Heydar Abad}, E.~{Ozfatura}, O.~{Ercetin}, and D.~{Gündüz}, ``Dynamic
  content updates in heterogeneous wireless networks,'' in \emph{2019 15th
  Annual Conference on Wireless On-demand Network Systems and Services (WONS)},
  2019, pp. 107--110.

\end{thebibliography}

\end{document}